\documentclass[sigconf,nonacm]{acmart}
\newcommand{\eg}{\emph{e.g.}}
\newcommand{\ie}{\emph{i.e.}}

\AtBeginDocument{%
  }

\usepackage{bm}
\usepackage{multirow}
\usepackage{colortbl}

\definecolor{hl}{RGB}{240,240,240}
\definecolor{bluesquare}{RGB}{78,149,217}
\def\red#1{\textcolor{red}{#1}}

\begin{document}

\title{Generalizing Video DeepFake Detection by Self-generated Audio-visual Pseudo-fakes}

\author{Zihe Wei}
\affiliation{%
  \institution{Ocean University of China}
  \city{Qingdao}
  \state{Shandong}
  \country{China}
}
\email{weizihe@stu.ouc.edu.cn}

\author{Yuezun Li}
\affiliation{%
  \institution{Ocean University of China}
  \city{Qingdao}
  \state{Shandong}
  \country{China}
}
\email{liyuezun@ouc.edu.cn}
\authornote{Corresponding author}

\renewcommand{\shortauthors}{Wei and Li}

\begin{abstract}
  Detecting video deepfakes has become increasingly urgent in recent years. Given the audio-visual information in videos, existing methods typically expose deepfakes by modeling cross-modal correspondence using specifically designed architectures with publicly available datasets. While they have shown promising results, their effectiveness often degrades in real-world scenarios, as the limited diversity of training datasets naturally restricts generalizability to unseen cases. To address this, we propose a simple yet effective method, called \textbf{AVPF}, which can notably enhance model generalizability by training with self-generated \textbf{A}udio-\textbf{V}isual \textbf{P}seudo-\textbf{F}akes. The key idea of AVPF is to create pseudo-fake training samples that contain diverse audio-visual correspondence patterns commonly observed in real-world deepfakes. We highlight that 
  AVPF is generated solely from authentic samples, and training relies only on authentic data and AVPF, without requiring any real deepfakes.
  Extensive experiments on multiple standard datasets demonstrate the strong generalizability of the proposed method, achieving an average performance improvement of up to $\bm{7.4}\%$. The code is available at \url{https://github.com/OUC-VAS/AVPF}.   
\end{abstract}    




\maketitle

\section{Introduction}



With the evolution of AI-based face generative techniques~\cite{idiff-boutros2023,facednerf-zhang2023,t23d-zhang2024}, video deepfakes have become increasingly realistic, leading to serious societal concerns, including privacy violations~\cite{sle-furizal2025}, financial fraud~\cite{bpc-vecchietti2025}, and even political manipulation~\cite{poldf-hameleers2024}. In response, a large body of research has focused on video deepfake detection~\cite{fa-cui2025,lvlm-yu2025,lsda-yan2024}, drawing growing attention and demonstrating promising effectiveness in safeguarding authenticity.

\begin{figure}[!t]
    \centering
    \includegraphics[width=\linewidth]{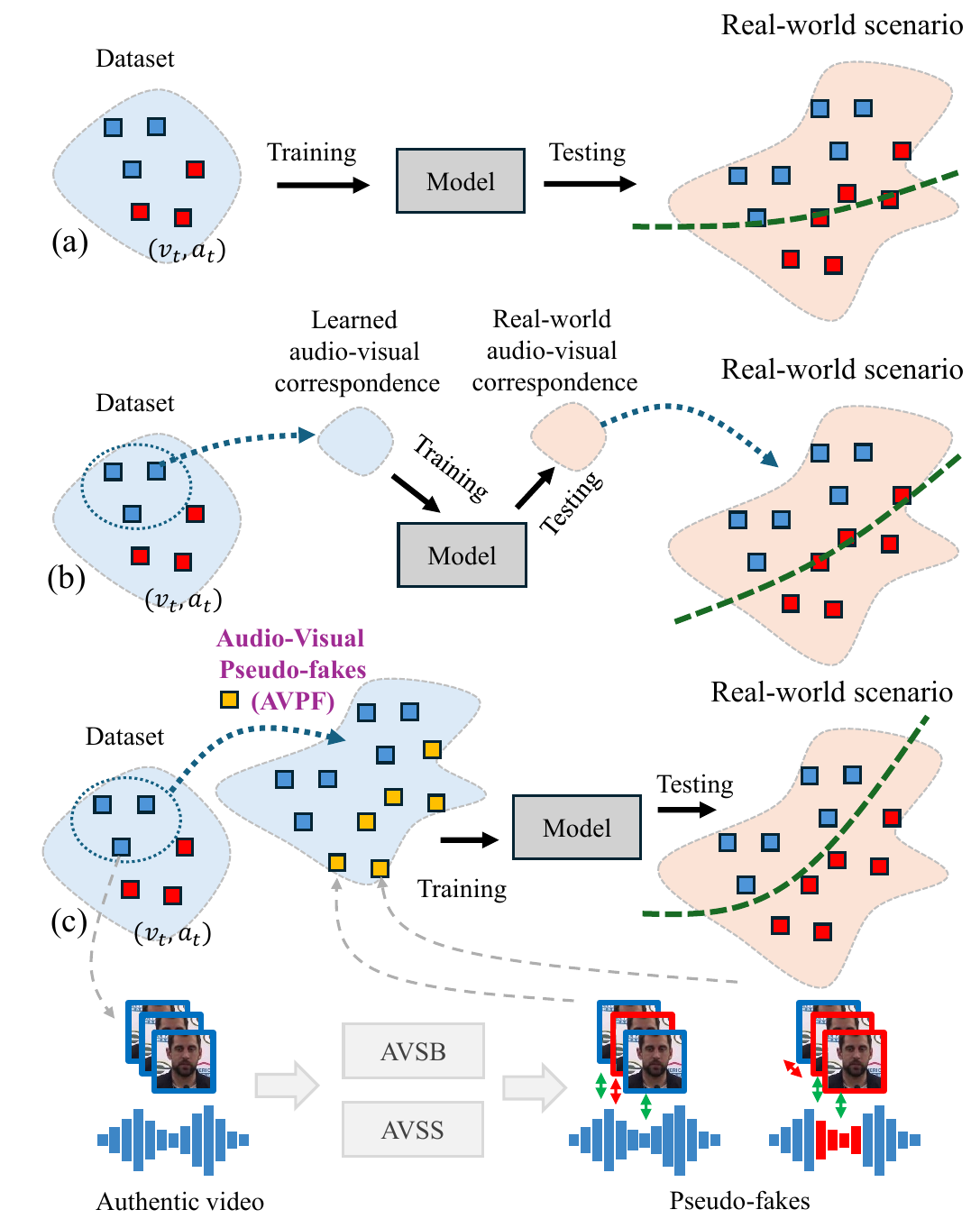}
    \vspace{-0.8cm}
    \caption{\small \textcolor{bluesquare}{$\blacksquare$} and \red{$\blacksquare$} denote authentic and deepfake training samples with visual $v_t$ and audio $a_t$ modality. (a) Training on known datasets suffers from limited generalizability, as the distribution of real-world deepfakes is far more complex. (b) Several recent works employ self-supervised learning to model the audio–visual correspondence of authentic videos and identify deepfakes by measuring deviations from the learned patterns. However, only relying on authentic samples struggles to cover the wide range of deepfake types in real-world scenarios. (c) In this paper, we propose self-generated Audio-Visual Pseudo-Fakes (AVPF), which effectively simulate the distribution of real-world deepfakes, leading to a notable improvement in generalizability. See text for more details.}
    \Description{A three-part comparison of conventional supervised deepfake detection, authentic-only self-supervised detection, and the proposed AVPF training strategy using authentic and generated pseudo-fake audio-visual samples.}
    \label{fig:overview}
    \vspace{-0.4cm} 
\end{figure}

Leveraging both audio and visual modalities has recently emerged as a promising solution for video deepfake detection~\cite{ics,mtavpl,avh}. These methods are motivated by the fact that the audio-visual correspondence is generally consistent in authentic videos but often becomes misaligned in deepfakes. Generally, existing works design models to extract modality-specific features and then capture their correspondence, either through holistic learning frameworks (\eg, neural networks~\cite{mtavpl,mrdf}) or explicit modeling strategies (\eg, contrastive learning~\cite{mds,vfd}). Note that they are typically trained on existing datasets, thus inherently limiting their generalizability to unseen scenarios. Noticing this point, several studies introduce self-supervised learning schemes that deliberately construct and learn specific audio-visual mismatches~\cite{avad,avh}. Despite their effectiveness, they remain restricted by the diversity of training data and struggle to cover the wide range of deepfake types in real-world scenarios. 

In this paper, as illustrated in Fig.~\ref{fig:overview}, we propose a simple yet effective method to improve detection generalizability with a novel self-generated Audio-Visual Pseudo-Fake strategy (\textbf{AVPF}). The key idea of AVPF is to create pseudo-fake video samples that simulate diverse audio-visual correspondence patterns commonly observed in real-world deepfakes, which are then used as negative training data with authentic videos as positive training data. By training on them, the model is driven to improve its understanding of generalizable representations. 

AVPF solely relies on authentic videos, considering two complementary aspects: \textbf{1) Inter-modality inconsistency}: it represents a temporal mismatch between the audio and the visual facial appearance, reflecting real-world scenarios where either the face or the audio is independently manipulated. To simulate this case, we propose the Audio-Visual Self-blending (\textbf{AVSB}) strategy, which temporally self-blends one modality within a random window, \ie, either the visual facial frames or the audio. \textbf{2) Intra-modality inconsistency}: it refers to temporal mismatch within each modality. In some deepfakes, audio–visual snippets are jointly manipulated, resulting in minimal inter-modality inconsistency. For such cases, it is more effective to focus on the manipulation traces introduced within individual modalities. To this end, we introduce Audio-visual Self-splicing (\textbf{AVSS}) strategy, which selects and splices appropriate audio-visual snippets from the same video into the desired location, deliberately introducing subtle artifacts within each modality.

While pseudo-fakes have been explored in visual-only deepfake detection~\cite{sbi,sbv}, \textbf{AVPF is uniquely valuable for two reasons:} First, extending pseudo-fakes from single-modal to multi-modal settings is non-trivial: existing methods consider only spatial blending artifacts within individual frames, whereas AVPF needs to simulate both spatial and temporal abnormal correlations across and within modalities that are commonly witnessed in real-world deepfakes. Second, AVPF is among the first works to investigate the feasibility of pseudo-fake generation in the multi-modal domain, offering a fresh perspective and establishing a baseline to inspire future research.


Our method is evaluated on multiple mainstream benchmarks and thoroughly compared with state-of-the-art methods. The experimental results show that AVPF improves the performance by up to $\bm{7.4}\%$ on average, demonstrating its efficacy in enhancing detection generalizability and truly aligning with our research philosophy: \textbf{Simplicity gets the job done}. In summary, our main contributions are threefold:
\begin{itemize}
    \item To the best of our knowledge, this work is among the first to incorporate pseudo-fake generation within multi-modality, providing a simple yet effective solution for video deepfake detection.
    \item To improve generalizability, we propose a self-generated Audio-Visual Pseudo-Fake (AVPF) strategy that models both inter- and intra-modality inconsistencies commonly observed in real-world deepfakes.
    \item Extensive experiments under various evaluation settings demonstrate the effectiveness of the proposed method, suggesting new possibilities for pseudo-fake-based strategies for multi-modal deepfake detection.
\end{itemize}
  
\section{Related Works}

\noindent\textbf{Audio-Visual Deepfake Detection.} 
Leveraging both audio and visual modalities for video deepfake detection has emerged as an active research direction in recent years. Compared to previous methods that rely on a single modality, \ie, purely visual deepfake detection~\cite{eddw,aunet,luo2024beyond} or audio deepfake detection~\cite{slim,zhangcan}, audio-visual methods exploit more comprehensive cues such as cross-modal correspondence. 
Complementary to existing single-modality methods, audio-visual deepfake detection has demonstrated its growing importance in the multimedia forensics field.

Particularly, the data-driven paradigm has drawn lots of attention, where task-specific architectures are designed and then trained using relevant datasets. For instance, a number of recent methods develop multi-modal feature extractors and automatically learn cross-modal inconsistencies, including MDS~\cite{mds}, VFD~\cite{vfd}, JAVDD~\cite{javdd}, LipFD~\cite{lipfd}, MRDF~\cite{mrdf}, MMMS-BA~\cite{mmms}, DiMoDif~\cite{dimodif}, AVGraph~\cite{avgpraph}, ART-AVDF~\cite{art-avdf}, BA-TFD+~\cite{batfd+}, and AV-LTI~\cite{astrid2025audio}.
Moreover, AVFF~\cite{avff} aligns audio and visual modalities and reconstructs audio and visual content by mixing cross-modal representations. A classifier is then trained to expose deepfakes by evaluating reconstruction quality.
Despite their promising results, these methods tend to overfit to dataset-specific biases, which limits their generalizability to unseen scenarios.

To mitigate this, self-learning strategies have been proposed that learn audio-visual correspondence solely on authentic videos from publicly available datasets, which detect deepfakes by measuring cross-modality similarity, \eg, AVAD~\cite{avad}, SpeechForensics~\cite{speechforensics}, and AVH-Align~\cite{avh}. 
By relying exclusively on authentic data, these methods naturally reduce dependence on manipulation-specific artifacts. 
Similar ideas can be found in one-class learning across various tasks~\cite{lee2024multi,narayan2024hyp,khalid2020oc,wang2022dku}.
However, relying solely on authentic data limits the exposure to diverse forgery patterns, preventing the model from fully unleashing its potential for generalized detection.

\smallskip
\noindent\textbf{Pseudo-fake Generation.} 
Pseudo-fake faces refer to intentionally generated facial images that contain artifacts similar to those introduced by deepfake techniques. In training, these faces are regarded as ``deepfake'' samples to improve generalizability. This strategy has been widely adopted in visual deepfake detection~\cite{li2019exposing,facexray,sbi,lsc,sladd,stc,vb,stsbv,sbv}. Typically, these methods create pseudo-fake facial images through introducing various blending artifacts in either self-blending or multi-source blending manners~\cite{li2019exposing,facexray,sbi,lsc,sladd}. More recently, this idea has been extended into video sequences, where both spatial and temporal deepfake-style artifacts are deliberately introduced. For example, the method in~\cite{vb} simulates facial feature drift via per-frame warping and self-blending, while the method in~\cite{stsbv} creates pseudo-fakes with multiple categories of spatial and temporal artifacts. It is important to note that the pseudo-fakes generated by these methods are solely within a single visual modality. In this paper, we take a step further and introduce a novel self-generated Audio-Visual Pseudo-Fake (AVPF) method dedicated to generalizing audio-visual deepfake detection. To the best of our knowledge, this is among the first attempts at pseudo-fake generation in a multi-modal setting.

\section{Proposed Method}
The proposed AVPF consists of two strategies, Audio-Visual Self-Blending (\textbf{AVSB}) strategy and Audio-Visual Self-Splicing (\textbf{AVSS}) strategy, which are designed to simulate the inter- and intra-modality inconsistency patterns commonly observed in deepfakes (see Figs.~\ref{fig:avsb} and \ref{fig:avss}). 
Notably, this is among the first attempts at multi-modal pseudo-fake generation, and AVPF relies solely on authentic videos.

\subsection{Audio-Visual Self-Blending Strategy}
\begin{figure}[!t]
    \centering
    \includegraphics[width=0.8\linewidth]{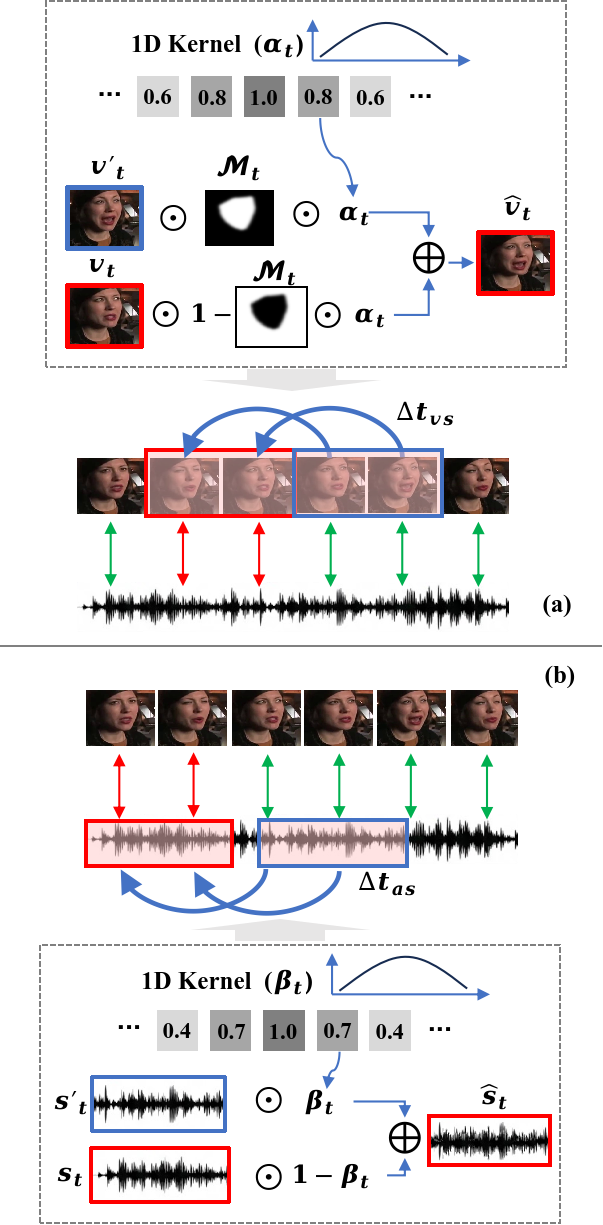}
    \vspace{-0.3cm}
    \caption{\small Overview of Audio-Visual Self-Blending (AVSB) strategy. (a) shows visual self-blending and (b) represents audio self-blending. The $\textcolor{green}{\bm{\updownarrow}}$ and $\textcolor{red}{\bm{\updownarrow}}$ symbols indicate consistent and inconsistent audio–visual correspondence, respectively. }
    \Description{Two AVSB pipelines: visual self-blending shifts and blends face frames while preserving audio, and audio self-blending shifts and blends audio while preserving the visual stream.}
    \label{fig:avsb}
     \vspace{-0.45cm}
\end{figure}

This strategy is proposed to simulate common deepfake scenarios in which forgery is applied to only one modality, \eg, manipulating facial expressions while leaving the audio unchanged, or altering the audio content while preserving the original facial appearance (see Fig.~\ref{fig:avsb}).

Formally, denote an authentic audio-visual video sample as $\bm{x}=\{(v_t,a_t)\}_{t=1}^{T}$, where $v_t$ denotes the $t$-th face frame and $a_t$ denotes the corresponding audio segment. Given such a video, we create a pseudo-fake sample by performing self-blending on either the audio or the visual modality, with the goal of intentionally introducing cross-modal inconsistency. Accordingly, this strategy yields two symmetric variants, \textit{visual self-blending} and \textit{audio self-blending}. The former perturbs only the visual modality while preserving the original audio, whereas the latter modifies only the audio stream while keeping the visual content unchanged.

\smallskip
\noindent\textbf{Visual Self-blending.}
Given the visual frames $\{v_t\}_{t=1}^{T}$ from video $\bm{x}$, a straightforward solution to self-blending is to temporally shift the visual track by $\Delta t_{vs} \in \{\pm 1, \ldots, \pm T\}$ and blend the shifted video with the original one using a factor $\alpha$. \textbf{While this operation can introduce cross-modal inconsistency, the resulting artifacts are overly pronounced and easily noticeable. Such intensive artifacts are unlikely to appear in high-quality deepfake videos}, thus making this approach ineffective for our purpose. As such, we design a more delicate self-blending process that intentionally introduces subtle and realistic artifacts, which are commonly observed in real-world deepfake scenarios.

Firstly, we temporally shift the video by $\Delta t_{vs}$ in either forward ($\Delta t_{vs} \geq 0$) or backward direction ($\Delta t_{vs} < 0$), resulting in a shifted video $\{v'_t\}_{t=1}^{T}$, where $v'_t = v_{t + \Delta t_{vs}}$. In practice, forgeries typically occur within localized temporal regions rather than across the entire video. Thus, we introduce a set of temporal window masks $\{p_1,...,p_n\}$, where each $p_i \in \{0,1\}^T$, to control the blending range. For example, $p_i = \{0\}^{t_1}_{j=1} \cup \{1\}^{t_2}_{j=t_1+1} \cup \{0\}^{T}_{j=t_2+1}$. The consecutive range of $[t_1 + 1,t_2]$ refers to window for blending.
To further mitigate boundary artifacts, we smooth the window boundaries using a 1D triangular smoothing kernel $h$. This process yields a time-varying blending factor $\alpha_t \in [0,1]$ for each frame, where $\alpha_t = 0$ indicates no blending and $\alpha_t = 1$ denotes full replacement by the shifted video.

Secondly, we restrict the blending operation to the facial region while leaving the background unchanged, since deepfake manipulations are predominantly localized to the facial region. To simulate this, we obtain a facial mask $\mathcal{M}_t$ by calculating a convex hull polygon over the detected facial landmarks. 
To enhance blending diversity, we apply an elastic deformation $\mathcal{T}$ to the mask, followed by Gaussian smoothing $g$ to alleviate artifacts near the blending boundaries. The final facial mask is thus obtained as $\mathcal{M}_t \leftarrow g * \mathcal{T}(\mathcal{M}_t) \in [0,1]^{w \times h \times 3}$, where $w$ and $h$ are the width and height of the frames, respectively.

Finally, inside a temporal window denoted by $p_i$, the blended frame at time $t$  is generated as 
\begin{equation}
    \hat{v}_t = v_t \cdot (1 - \mathcal{M}_t \cdot \alpha_t) + v'_t \cdot \mathcal{M}_t \cdot \alpha_t.
\end{equation}

\smallskip\noindent\textbf{Audio Self-blending.}
For the audio $\{a_t \}_{t=1}^{T}$ from video $\bm{x}$, we first convert it into Mel-spectrogram representation $\{s_t \}_{t=1}^{T}$, where $s_t$ denotes the Mel-frequency bins at time step $t$. Analogous to visual self-blending, we temporally shift the Mel-spectrogram by $\Delta t_{as}$ to obtain a shifted Mel-spectrogram $\{s'_t \}_{t=1}^{T}$, where $s'_t = s_{t + \Delta t_{as}}$. The shifted and original Mel-spectrograms are then blended using the same strategy as in visual self-blending. Specifically, a set of temporal window masks is constructed and smoothed using a one-dimensional triangular kernel to produce a time-varying blending factor $\beta_t \in [0,1]$, which controls the degree of audio blending at each time step. The blended Mel-spectrogram at time $t$ can be written as
\begin{equation}
    \hat{s}_t = s_t \cdot (1 - \beta_t) + s'_t \cdot \beta_t.
\end{equation}
Then we invert the blended Mel-spectrogram $\{\hat{s}_t \}_{t=1}^{T}$ back to audio $\{\hat{a}_t\}_{t=1}^{T}$.

\subsection{Audio-Visual Self-Splicing Strategy}
\begin{figure}[!t]
    \centering
    \includegraphics[width=0.8\linewidth]{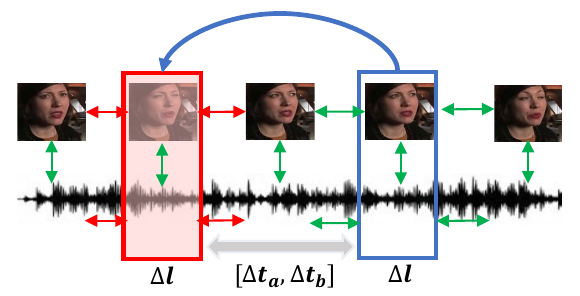}
    \vspace{-0.4cm}
    \caption{\small Overview of Audio-Visual Self-Splicing (AVSS) strategy. The $\textcolor{green}{\bm{\leftrightarrow}}$ and $\textcolor{red}{\bm{\leftrightarrow}}$ symbols indicate intra-modal consistency and inconsistency, respectively. }
    \Description{The AVSS pipeline selects a visually similar but temporally distant audio-visual snippet and splices it into the target interval to introduce subtle intra-modal temporal inconsistency.}
    \label{fig:avss}
    \vspace{-0.2cm} 
\end{figure}

This strategy is designed to simulate deepfake scenarios involving joint manipulation of both visual and audio modalities, in which cross-modal inconsistency is minimal but the forgery can still be revealed by inspecting modality-specific temporal inconsistencies. For example, lip movements and the corresponding audio are manipulated together within a localized temporal window (see Fig.~\ref{fig:avss}).

An intuitive approach is to directly replace a snippet with another one randomly selected from the same video. \textbf{However, such an operation can introduce clear intra-modality inconsistency, which rarely exists in real-world deepfakes.} For better simulation, we intentionally reduce the artifacts by splicing similar snippets. Specifically, for a short snippet $\{(v_t,a_t)\}_{t=i}^{i+ \Delta l}$, we find another most similar snippet with the same length in a temporal range of $[\Delta t_{a}, \Delta t_{b}]$. Suppose that $\{(v'_t,a'_t)\}_{t=i'}^{i'+ \Delta l}$ is the found snippet, in which $\Delta t_{a} \leq |i - i'| \leq \Delta t_{b}$.

To find the similar snippet, we simply measure the mean absolute difference of frames, \ie, $\sum_{i,i'}|v_i - v'_{i'}|$, and pick the snippet that has minimal difference. 
Finally, we obtain the self-spliced video as $\{(v_t,a_t)\}_{t=1}^{i-1} \cup \{(v'_t,a'_t)\}_{t=i'}^{i'+ \Delta l} \cup \{(v_t,a_t)\}_{t=i+\Delta l + 1}^{T}$ .

\section{Experiments}

\begin{table*}[!t]
  \caption{\small The detection performance of our method compared to other state-of-the-arts. Note that \textbf{RF} means using both real and fake data, while \textbf{R} indicates using real data only. \textbf{Cross eval.} denotes that the model is trained on one dataset but evaluated on other datasets. \textbf{Trim test} indicates the leading silent segment at the beginning of each test video is trimmed, as identified in~\cite{avh}. $(\dagger)$ and $(*)$ denote retraining models and using provided weights, respectively.} 
  \vspace{-0.4cm}
  \label{table:compare}
  \small
  \centering
    \begin{tabular}{lccccccccccccc}
      \toprule
      \multirow{2}{*}{Method} & \multirow{2}{*}{Venue} & Train & Used & Cross & Trim & \multicolumn{2}{c}{FAVC} & \multicolumn{2}{c}{AV1M} & \multicolumn{2}{c}{AVLips}& \multicolumn{2}{c}{Average} \\
      \cmidrule(lr){7-8} \cmidrule(lr){9-10} \cmidrule(lr){11-12} \cmidrule(lr){13-14}
      & & set & data & eval. & test & AUC & AP & AUC & AP & AUC & AP & AUC & AP \\
      \midrule
      MDS \cite{mds} & MM'20 & FAVC & RF & $\times$ & $\times$ & 51.9 & - & - & - & - & - & - & - \\
      JAVDD \cite{javdd} & ICCV'21 & FAVC & RF & $\times$ & $\times$ & 52.0 & - & - & - & - & - & - & - \\
      VFD \cite{vfd} & TOMM'23 & FAVC & RF & $\times$ & $\times$ & 86.1 & - & - & - & - & - & - & - \\
      DiMoDif \cite{dimodif} & arxiv'24 & FAVC & RF & $\times$ & $\times$ & 99.7 & - & - & - & - & - & - & - \\
      DiMoDif \cite{dimodif} & arxiv'24 & AV1M & RF & $\times$ & $\times$ & - & - & 96.3 & - & - & - & - & - \\
      MRDF-CE \cite{mrdf} & ICASSP'24 & FAVC & RF & $\times$ & $\times$ & 98.2 & 99.6 & - & - & - & - & - & - \\
      MRDF-CE \cite{mrdf} & ICASSP'24 & AV1M & RF & $\times$ & $\times$ & - & - & 97.8 & 97.2 & - & - & - & - \\
      MMMS-BA \cite{mmms} & IJCB'24 & FAVC & RF & $\times$ & $\times$ & 98.9 & - & - & - & - & - & - & - \\
      AVFF \cite{avff} & CVPR'24 & FAVC & RF & $\times$ & $\times$ & 99.1 & - & - & - & - & - & - & - \\
      AVGraph \cite{avgpraph} & IJCV'24 & FAVC & RF & $\times$ & $\times$ & 99.9 & - & - & - & - & - & - & - \\
      MAV-PL \cite{mtavpl} & AAAI'25 & FAVC & RF & $\times$ & $\times$ & 100.0 & - & - & - & - & - & - & - \\
      AVH-Align/sup \cite{avh} & CVPR'25 & FAVC & RF & $\times$ & $\times$ & 99.2 & 100.0 & - & - & - & - & - & - \\
      AVH-Align/sup \cite{avh} & CVPR'25 & AV1M & RF & $\times$ & $\times$ & - & - & 100.0 & 100.0 & - & - & - & - \\
      \midrule
      MRDF-CE$^{\dagger}$ \cite{mrdf} & ICASSP'24 & FAVC & RF & $\times$ & $\checkmark$ & 86.2 & 14.7 & - & - & - & - & - & - \\
      MRDF-CE$^{\dagger}$ \cite{mrdf}& ICASSP'24 & AV1M & RF & $\times$ & $\checkmark$ & - & - & 50.1 & 21.6 & - & - & - & - \\
      AVH-Align/sup \cite{avh} & CVPR'25 & FAVC & RF & $\times$ & $\checkmark$ & 99.2 & 100.0 & - & - & - & - & - & - \\
      AVH-Align/sup \cite{avh} & CVPR'25 & AV1M & RF & $\times$ & $\checkmark$ & - & - & 83.1 & 94.4 & - & - & - & - \\
      AVH-Align/sup$^{*}$ \cite{avh} & CVPR'25 & FAVC & RF & $\times$ & $\checkmark$ & 99.1 & 100.0 & - & - & - & - & - & - \\
      AVH-Align/sup$^{*}$ \cite{avh} & CVPR'25 & AV1M & RF & $\times$ & $\checkmark$ & - & - & 83.0 & 94.4 & - & - & - & - \\
      \midrule
      DiMoDif \cite{dimodif} & arxiv'24 & FAVC & RF & $\checkmark$ & $\times$ & - & - & 54.0 & 77.9 & - & - & - & - \\
      DiMoDif \cite{dimodif} & arxiv'24 & AV1M & RF & $\checkmark$ & $\times$ & 90.7 & 99.7 & - & - & - & - & - & - \\
      MMMS-BA \cite{mmms} & IJCB'24 & FAVC & RF & $\checkmark$ & $\times$ &  - & - & 90.9 & - & - & - & - & - \\
      MMMS-BA \cite{mmms} & IJCB'24 & AV1M & RF & $\checkmark$ & $\times$ &  95.5 & - & - & - & - & - & - & - \\
      \midrule
       BA-TFD+$^{*}$ \cite{batfd+} & CVIU'23 & LAV-DF & RF & $\checkmark$ & $\checkmark$ & 64.9 & 98.9 & 62.9 & 85.2 & 48.3 & 56.1 & 58.7 & 80.1 \\
       MRDF-CE$^{\dagger}$ \cite{mrdf} & ICASSP'24 & AV1M & RF & $\checkmark$ & $\checkmark$ & 52.8 & 2.6 & - & - & 49.7 & 43.0 & - & - \\
       MRDF-CE$^{\dagger}$ \cite{mrdf} & ICASSP'24 & FAVC & RF & $\checkmark$ & $\checkmark$ & - & - & 50.2 & 24.3 & 65.9 & 69.2 & - & - \\
       LipFD$^{*}$ \cite{lipfd} & NeurIPS'24 & AVLips & RF & $\checkmark$ & $\checkmark$ & 85.5 & 99.6 & 50.7 & 75.7 & - & - & - & - \\
       AVH-Align/sup \cite{avh} & CVPR'25 & AV1M & RF & $\checkmark$ & $\checkmark$ & 70.8 & 99.1 & - & - & - & - & - & - \\
       AVH-Align/sup \cite{avh} & CVPR'25 & FAVC & RF & $\checkmark$ & $\checkmark$ & - & - & 63.6 & 82.2 & - & - & - & - \\
       AVH-Align/sup$^{*}$ \cite{avh} & CVPR'25 & AV1M & RF & $\checkmark$ & $\checkmark$ & 71.2 & 99.1 & - & - & 48.6 & 55.6 & - & - \\
       AVH-Align/sup$^{*}$ \cite{avh} & CVPR'25 & FAVC & RF & $\checkmark$ & $\checkmark$ & - & - & 63.5 & 82.2 & 96.1 & 94.1 & - & - \\
      \midrule
      AVAD \cite{avad} & CVPR'23 & LRS & R & $\checkmark$ & $\checkmark$ & 84.7 & 99.5 & 54.3 & 76.3 & 73.2$^{*}$ & 77.1$^{*}$ & 70.7 & 84.3 \\
      SpeechForensics \cite{speechforensics} & NeurIPS'24 & VoxCeleb2 & R & $\checkmark$ & $\checkmark$ & \textbf{98.8} & \textbf{100.0} & 68.2 & 83.5 & \underline{92.4}$^{*}$ & \underline{94.9}$^{*}$ & 86.5 & \underline{92.8} \\
      AVH-Align \cite{avh} & CVPR'25 & VoxCeleb2 & R & $\checkmark$ & $\checkmark$ & 94.6 & 99.8 & \underline{83.5} & \underline{93.5} & 86.6$^{*}$ & 76.8$^{*}$ & \underline{88.2} & 90.0\\
      \rowcolor{hl} \textbf{AVPF} & \textbf{Ours} & VoxCeleb2 & R & $\checkmark$ & $\checkmark$ & \underline{97.8} & \underline{99.9} & \textbf{89.2} & \textbf{96.2} & \textbf{97.6} & \textbf{97.8} & \textbf{94.9} & \textbf{98.0} \\
      \bottomrule
    \end{tabular}
\end{table*}

\subsection{Experimental Settings}\label{sec:setting}

\smallskip\noindent\textbf{Datasets and metrics.}
Our method is evaluated on four standard audio-visual datasets: FakeAVCeleb~\cite{favc}, AV-Deepfake1M~\cite{av1m}, AVLips~\cite{lipfd} and TalkingHeadBench~\cite{talkingheadbench}.
Since our method only relies on real samples, we follow the evaluation settings in~\cite{avh}, where the model is trained using 50K samples from the VoxCeleb2 dataset and directly evaluated on the test set of FakeAVCeleb as well as on 10K samples from the validation set of AV-Deepfake1M. 
In addition, our method is evaluated on the entire AVLips dataset~\cite{lipfd}.
To evaluate the performance of our method on diffusion-based forgeries, we further conduct experiments on the TalkingHeadBench~\cite{talkingheadbench}.
Following~\cite{avh}, detection performance is reported using the Area Under the Curve (AUC) and Average Precision (AP) metrics.

\smallskip\noindent\textbf{Backbone.}
We adopt AV-HuBERT~\cite{avhubert} to extract features from both audio and visual modalities. AV-HuBERT is a self-supervised audio–visual representation learning framework that jointly models speech and lip movements to learn robust, synchronized audio–visual features. This model has also been adopted in related work~\cite{avh}.


\smallskip\noindent\textbf{Implementation Details.}
Our method is implemented using PyTorch 2.2.0~\cite{pytorch} with a single NVIDIA RTX 3090 GPU. 
During preprocessing, we follow~\cite{avh} that extracts $96 \times 96$ lip ROI sequences from videos and the corresponding audio waveforms using FFmpeg.
In Visual Self-blending, we set window shift $\Delta t_{vs}= 2$ and length randomly sampled from $[0.5, 1.5]$ seconds. In {Audio Self-blending}, the shift $\Delta t_{as}$ is randomly sampled from $[0.02, 0.05]$ seconds. Both blending strategies allow bidirectional shifts. 
In {Audio-Visual Self-Splicing Strategy}, the default spliced snippet length is set to $\Delta l = 1$ frame and the offset is set to $[\Delta t_a = 0.5, \Delta t_b=1]$.   
In training, the backbone model is trained using a simple binary classification objective, with the Adam optimizer using a learning rate of $1\times10^{-3}$.

\smallskip\subsection{Comparison with State-of-the-arts}\label{sec:Comparison}
Our method is compared with several state-of-the-art methods, including MDS~\cite{mds}, JAVDD~\cite{javdd}, VFD~\cite{vfd}, DiMoDif~\cite{dimodif}, MRDF-CE
~\cite{mrdf}, MMMS-BA~\cite{mmms}, AVFF~\cite{avff}, AVGraph~\cite{avgpraph}, MAV-PL~\cite{mtavpl}, AVH-Align~\cite{avh}, BA-TFD+~\cite{batfd+}, LipFD~\cite{lipfd}, AVAD~\cite{avad}, SpeechForensics~\cite{speechforensics}.

Table~\ref{table:compare} shows the comparison results of our method with state-of-the-arts. For the first part, the performance of MDS, JAVDD, MRDF-CE, AVFF, AVGraph, and MAV-PL is taken from the MAV-PL paper, while results for the remaining methods are obtained from their respective papers. For the second part, we retrain MRDF-CE using its released code, and directly use the provided weights for AVH-Align. The performance in the third part is all taken from original papers. In the fourth part, BA-TFD+, LipFD, and AVH-Align are evaluated using their provided weights, whereas MRDF-CE is retrained. For the fifth part, we report the performance of all methods on the FAVC and AV1M datasets as reported in AVH-Align, and evaluate them on the AVLips dataset using the provided weights. 
We can observe that most methods require both real and fake training data. Due to dataset bias, these methods often exhibit favorable, and sometimes even near-perfect, performance in intra-dataset evaluations, but show limited generalizability across datasets. In contrast, AVAD, SpeechForensics, and AVH-Align can be trained solely on real data by self-supervised strategies, showing notable improvements in generalizability. However, as the distribution of real-world data alone cannot encompass the myriad of forgery techniques, the model's potential for robust generalization remains constrained. In comparison, our method (AVPF) largely increases the training diversity by using the newly proposed Audio-Visual Pseudo-Fakes, enabling the model to more deeply learn audio–visual correspondence in video deepfakes and thereby further improving generalizability. Quantitatively, compared to the most recent method AVH-Align, our method achieves performance gains of $6.7\%$ in AUC and  $8.0\%$ in AP, respectively.

\begin{table}[!t]
  \caption{\small Comparison with AVH-Align on TalkingHeadBench.}
  \vspace{-0.4cm}
  \centering
  \small
  \label{tab:thb_comp}
  \begin{tabular}{lcc}
    \toprule
    Method & AUC & AP \\
    \midrule
    AVH-Align~\cite{avh} & 28.5 & 44.9 \\
    \rowcolor{hl} \textbf{AVPF (Ours)}      & \textbf{77.8} & \textbf{79.1} \\
    \bottomrule
  \end{tabular}
  \vspace{-0.4cm}  
\end{table}

Moreover, we compare our method with the latest method, AVH-Align, on the TalkingHeadBench dataset. As shown in Table~\ref{tab:thb_comp},
AVH-Align shows a substantial performance drop, achieving only $28.5\%$ AUC and $44.9\%$ AP, while 
our method reaches $77.8\%$ AUC and $79.1\%$ AP. This is because in the TalkingHeadBench dataset, the audio-visual inconsistency is less noticeable. Since AVH-Align relies on heavy audio-visual misalignment, it struggles to capture subtle traces.
In contrast, thanks to the curation in AVSB and AVSS, AVPF can generate pseudo-fakes with subtle audio-visual correspondence, enabling the model to capture fine-grained traces present in the TalkingHeadBench dataset.

\subsection{Ablations}

\smallskip\noindent\textbf{Effect of Each Proposed Strategy.}
Table~\ref{table:ablation1} reports the ablation results obtained by disabling the proposed AVSB and AVSS strategies. As shown, while individual datasets may exhibit slight variations, the removal of any augmentation strategy leads to a consistent decline in average performance. This trend underscores the overall contribution of each proposed strategy to the model's effectiveness.
Among them, Visual Self-Blending (VSB) plays a crucial role in generalization. Disabling VSB causes a notable drop on all datasets (Average: 81.2/92.4 vs.\ 94.9/98.0), indicating that the model without VSB tends to overfit to unimportant temporal patterns.
Discarding AVSS also leads to a noticeable drop, mainly on AV1M (79.5/92.1), yielding a substantially lower average (90.9/95.8).
In contrast, disabling Audio Self-Blending (ASB) results in a relatively small decrease in the averaged score (94.2/97.8), suggesting that ASB provides an additional but secondary effect.
With all components enabled, AVPF achieves the best overall performance, reaching 94.9/98.0 on average.

\begin{table}[htbp]
\vspace{-0.1cm}
  \caption{\small Ablation studies on the effect of each proposed strategy.}
  \vspace{-0.4cm}
  \centering
  \small
  \label{table:ablation1}
  \resizebox{\linewidth}{!}{
  \begin{tabular}{l cc cc cc cc}
    \toprule
    \multirow{2}{*}{Method} &
    \multicolumn{2}{c}{FAVC} &
    \multicolumn{2}{c}{AV1M} &
    \multicolumn{2}{c}{AVLips} &
    \multicolumn{2}{c}{Average} \\
    \cmidrule(lr){2-3}\cmidrule(lr){4-5}\cmidrule(lr){6-7} \cmidrule(lr){8-9}
    & AUC & AP & AUC & AP & AUC & AP & AUC & AP \\
    \midrule
    w/o VSB &  81.7 & 99.5 & 84.3 & 94.6 & 77.5 & 83.2 & 81.2 & 92.4\\
    w/o ASB & 97.1 & \textbf{99.9} & 87.7 & 95.7 & \textbf{97.7} & \textbf{97.9} & 94.2 & 97.8\\
    w/o AVSS & \textbf{97.9} & \textbf{99.9} & 79.5 & 92.1 & 95.4 & 95.3 & 90.9 & 95.8\\
    AVPF (ours) & 97.8 & \textbf{99.9} & \textbf{89.2} & \textbf{96.2} & 97.6 & 97.8 & \textbf{94.9} & \textbf{98.0}\\
    \bottomrule
  \end{tabular}
  }
\vspace{-0.1cm}
\end{table}

\smallskip\noindent\textbf{Analysis on Audio-Visual Self-Blending (AVSB).}
As summarized in Table~\ref{tab:ablation-avsb} (A), the shift length $\Delta t_{vs}$ in VSB achieves the best trade-off when shifting by 2 frames, surpassing both milder (1 frame) and more aggressive (3 frames) perturbations.
For shift length $\Delta t_{as}$ in ASB, the results in Table~\ref{tab:ablation-avsb} (B) indicate that the 0.02--0.05 second interval is most effective. Performance drops with shorter (0.01--0.02 seconds) or longer (0.05--0.10 seconds) shifts. This trend suggests that overly small shifts leave weak forgery traces, whereas excessive shifts may disrupt informative temporal dynamics. 
Table~\ref{tab:ablation-avsb} (C) shows the effect of localized temporal windows. Compared with applying self-blending over the entire sequence (\ie, $p=\{1\}^T$), restricting augmentations to localized temporal windows (\ie, $\{p_1,\ldots,p_n\}$) further improves performance, achieving the best overall results (94.9/98.0). This observation highlights the importance of explicitly controlling when augmentations are applied, as it mitigates over-perturbation while preserving informative temporal cues.

\begin{table}[h]
  \caption{\small Ablation studies on various settings in AVSB.}
  \vspace{-0.4cm}
  \label{tab:ablation-avsb}
  \centering
  \small
  \resizebox{\linewidth}{!}{
  \begin{tabular}{l cc cc cc cc}
    \toprule
    \multirow{2}{*}{Setting} & 
    \multicolumn{2}{c}{FAVC} &
    \multicolumn{2}{c}{AV1M} &
    \multicolumn{2}{c}{AVLips} &
    \multicolumn{2}{c}{Average} \\
    \cmidrule(lr){2-3}\cmidrule(lr){4-5}\cmidrule(lr){6-7}\cmidrule(lr){8-9}
    & AUC & AP & AUC & AP & AUC & AP & AUC & AP \\
    \midrule

    \multicolumn{9}{l}{(A) Shift $\Delta t_{vs}$ in VSB} \\
    \midrule
    1 frame  & 96.7 & \textbf{99.9} & 85.4 & 94.9 & 96.0 & 96.4 & 92.7 & 97.1 \\
    2 frames & \textbf{97.8} & \textbf{99.9} & \textbf{89.2} & \textbf{96.2} & \textbf{97.6} & \textbf{97.8} & \textbf{94.9} & \textbf{98.0} \\
    3 frames & 93.8 & 99.8 & 87.4 & 95.7 & 96.6 & 97.0 & 92.6 & 97.5 \\
    \midrule

    \multicolumn{9}{l}{(B) Shift $\Delta t_{as}$ in ASB} \\
    \midrule
    0.01--0.02s & 96.8 & \textbf{99.9} & 87.3 & 95.6 & 96.8 & 97.1 & 93.6 & 97.5 \\
    0.02--0.05s & \textbf{97.8} & \textbf{99.9} & \textbf{89.2} & \textbf{96.2} & \textbf{97.6} & \textbf{97.8} & \textbf{94.9} & \textbf{98.0} \\
    0.05--0.10s & 95.0 & 99.8 & 85.3 & 94.8 & 93.5 & 92.3 & 91.3 & 95.6 \\
    \midrule

    \multicolumn{9}{l}{(C) Effect of localized temporal windows in AVSB} \\
    \midrule
    Entire $p=\{1\}^T$  & 95.3 & \textbf{99.9} & 84.2 & 94.5 & 96.2 & 96.4 & 91.9 & 96.9 \\
    Localized $\{p_1,...,p_n \}$  & \textbf{97.8} & \textbf{99.9} & \textbf{89.2} & \textbf{96.2} & \textbf{97.6} & \textbf{97.8} & \textbf{94.9} & \textbf{98.0} \\
    \bottomrule
  \end{tabular}
  }
\end{table}

\smallskip\noindent\textbf{Analysis on Audio-Visual Self-Splicing (AVSS).}
This part further analyzes the design choices of Audio-Visual Self-Splicing (AVSS).
As shown in Table~\ref{tab:ablation-avss} (A), a finding range of 0.5--1.0 seconds yields the best averaged performance, while smaller (0.2--0.5 seconds) or larger (1.0--1.5 seconds) ranges are slightly less effective. This result reflects a trade-off between perturbation realism and detection difficulty.
In addition, enabling similarity-based selection for spliced snippets consistently improves the average score (94.9/98.0 vs.\ 93.3/97.3), as reported in Table~\ref{tab:ablation-avss} (B). This implies that appearance-constrained copying introduces perturbations that are both challenging and semantically plausible.
Finally, the comparison in Table~\ref{tab:ablation-avss} (C) suggests that using 1 frame spliced length achieves the optimal performance. Increasing the number of frames generally degrades performance, likely due to the amplified temporal inconsistency caused by larger edits.

\begin{table}[h]
  \vspace{-2pt}
  \caption{\small Ablation studies on various settings in AVSS.}
  \vspace{-0.4cm}
  \label{tab:ablation-avss}
  \centering
  \small
  \resizebox{\linewidth}{!}{
  \begin{tabular}{l cc cc cc cc}
    \toprule
    \multirow{2}{*}{Setting} & 
    \multicolumn{2}{c}{FAVC} &
    \multicolumn{2}{c}{AV1M} &
    \multicolumn{2}{c}{AVLips} &
    \multicolumn{2}{c}{Average} \\
    \cmidrule(lr){2-3}\cmidrule(lr){4-5}\cmidrule(lr){6-7}\cmidrule(lr){8-9}
    & AUC & AP & AUC & AP & AUC & AP & AUC & AP \\
    \midrule
    \multicolumn{9}{l}{(A) Range $\Delta t_{a}-\Delta t_{b}$ in AVSS} \\
    \midrule
    0.2--0.5s & 96.6 & \textbf{99.9} & 87.5 & 95.7 & 97.1 & 97.3 & 93.7 & 97.6 \\
    0.5--1.0s & \textbf{97.8} & \textbf{99.9} & \textbf{89.2} & \textbf{96.2} & \textbf{97.6} & \textbf{97.8} & \textbf{94.9} & \textbf{98.0} \\
    1.0--1.5s & 96.3 & \textbf{99.9} & 87.5 & 95.7 & 96.1 & 96.2 & 93.3 & 97.3 \\
    \midrule

    \multicolumn{9}{l}{(B) Effect of similarity-based selection for spliced snippets in AVSS} \\
    \midrule
    w/o  & 96.0 & 99.8 & 87.9 & 95.5 & 95.9 & 96.5 & 93.3 & 97.3 \\
    w/   & \textbf{97.8} & \textbf{99.9} & \textbf{89.2} & \textbf{96.2} & \textbf{97.6} & \textbf{97.8} & \textbf{94.9} & \textbf{98.0} \\
    \midrule

    \multicolumn{9}{l}{(C) Number of spliced frames $\Delta l$ in AVSS} \\
    \midrule
    1 frame  & \textbf{97.8} & \textbf{99.9} & \textbf{89.2} & 96.2 & \textbf{97.6} & \textbf{97.8} & \textbf{94.9} & \textbf{98.0} \\
    2 frames & 96.0 & 99.8 & 87.7 & 95.5 & 95.9 & 95.7 & 93.2 & 97.0 \\
    3 frames & 96.8 & \textbf{99.9} & \textbf{89.2} & \textbf{96.3} & 96.2 & 96.6 & 94.1 & 97.6 \\
    
    \bottomrule
  \end{tabular}
  }
  \vspace{-2pt}
\end{table}

\smallskip\noindent\textbf{Training with Both Real and Fake Samples.}
We further investigate whether our method can enhance the supervised
learning when both real and fake data are available. For
a fair comparison, we apply the same Trim protocol to
both training and testing videos, removing the leading silent
segments as described in Sec.~\ref{sec:Comparison}. For FAVC, since the
number of real videos is limited (only 500), we augment
the real split by sampling additional real videos from its
original real source, VoxCeleb2. Specifically, for each of
the 500 real videos, we randomly retrieve 19 more real
videos while encouraging broader coverage of speakers and
scenes, resulting in approximately 10K real training videos
in total. The evaluation on the test sets strictly follows our
standard protocol in Sec.~\ref{sec:setting}. Note that both the baseline
method AVH-Align and our method are evaluated under
identical settings.

As reported in Table~\ref{tab:realfake}, integrating our method into AVH-Align achieves consistent improvements. Notably, when trained on FAVC and tested on AV1M, our method brings a substantial gain of $\bm{+15.5\%}$ in AUC and $\bm{+9.2\%}$ in AP. Similar trends are observed for other cross-dataset evaluations, indicating that the proposed pseudo-fake data effectively mitigates overfitting to dataset-specific biases.


\begin{table*}[!t]
  \caption{\small The detection performance of training with both \textbf{real and fake} data.}
  \vspace{-0.4cm}
  \label{tab:realfake}
  \centering
  \small
  \begin{tabular}{l c cc cc cc cc}
    \toprule
    \multirow{2}{*}{Method} &
    Train &
    \multicolumn{2}{c}{AV1M} &
    \multicolumn{2}{c}{FAVC} &
    \multicolumn{2}{c}{AVLips} &
    \multicolumn{2}{c}{Average} \\
    \cmidrule(lr){3-4}\cmidrule(lr){5-6}\cmidrule(lr){7-8} \cmidrule(lr){9-10}
    & set & AUC & AP & AUC & AP & AUC & AP & AUC & AP \\
    \midrule
    AVH-Align & \multirow{2}{*}{FAVC} & 57.7 & 80.3 & \textbf{99.9} & \textbf{100.0} & 98.3 & 98.2 & 85.3 & 92.8\\
    AVH-Align + \textbf{AVPF} & & \textbf{73.2} & \textbf{89.5} & 99.5 & \textbf{100.0} & \textbf{98.6} & \textbf{98.5} & \textbf{90.4} & \textbf{96.0}\\
    \midrule
    AVH-Align & \multirow{2}{*}{AV1M} & \textbf{99.2} & \textbf{99.7} & \textbf{94.7} & 99.8 & 87.9 & 85.0 & 93.9 & 94.8\\
    AVH-Align + \textbf{AVPF} & & 98.1 & 99.4 & \textbf{94.7} & \textbf{99.9} & \textbf{93.6} & \textbf{93.6} & \textbf{95.5} & \textbf{97.6} \\
    \bottomrule
  \end{tabular}
\end{table*}

\begin{figure*}[!t]
    \centering
    \includegraphics[width=\linewidth]{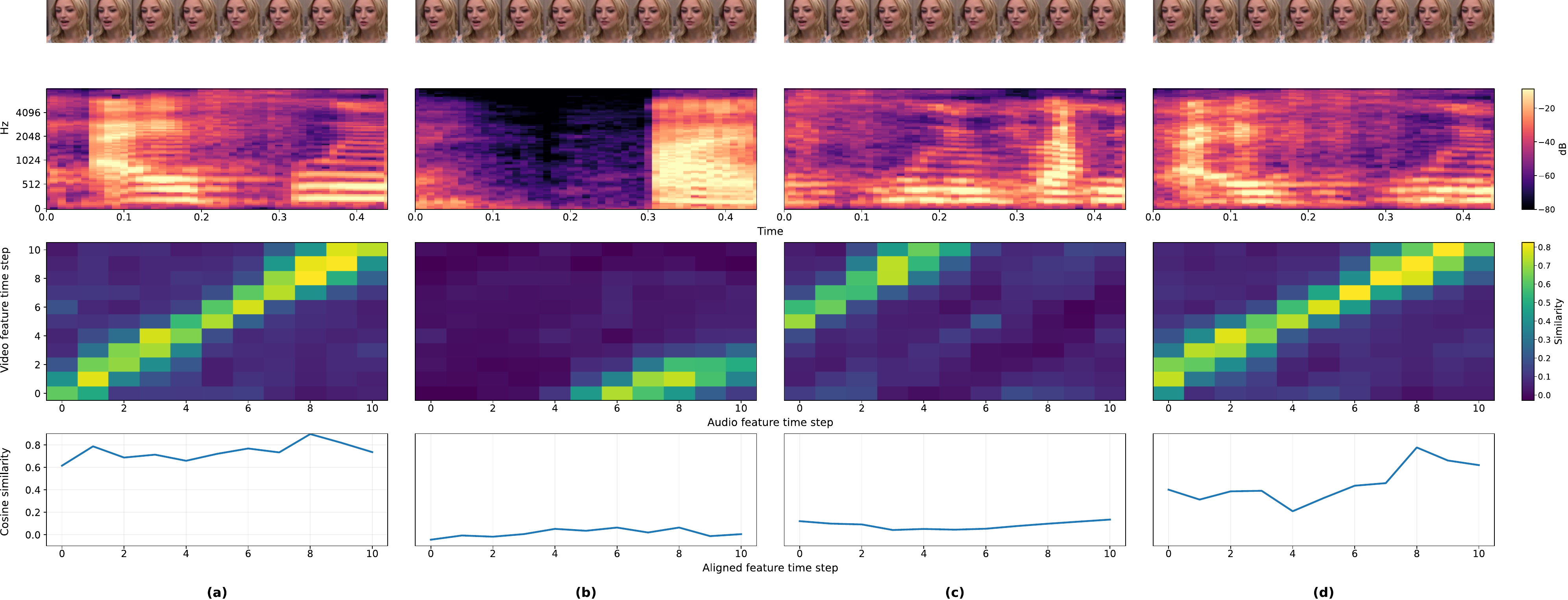}
    \vspace{-0.8cm}
    \caption{\small (a,b,c,d) correspond to analysis results of authentic videos, easy deepfakes, naive pseudo-fakes, and our method. The top-to-bottom rows show the video frames, the corresponding audio Mel-spectrograms, the local audio-video similarity matrix, and the frame-wise audio-video cosine similarity. It can be seen that our method could generate hard samples that contain subtle audio-visual inconsistency compared to easy deepfake and naive pseudo-fake. See text for more details.}
    \Description{Four columns compare authentic video, easy deepfake, naive pseudo-fake, and AVPF samples using frames, Mel-spectrograms, local audio-video similarity matrices, and frame-wise cosine similarity curves.}
    \label{fig:four_panel}
\end{figure*}

\subsection{Further Analysis}

\vspace{-4pt}
\smallskip\noindent\textbf{Does AVPF Really Simulate the Audio-visual Correspondence in Deepfakes?}
To validate whether our method effectively simulates the real audio-visual correspondence in deepfakes, we conduct the analysis presented in Fig.~\ref{fig:four_panel}. As shown, columns (a)-(d) correspond to authentic videos, easy deepfakes, naive pseudo-fakes, and our method, respectively. From top to bottom, the rows display the video frames, the corresponding audio Mel-spectrograms, the local audio-visual similarity matrix (LAVSM), and the frame-wise audio-visual similarity along the diagonal of the LAVSM (AVSD). Specifically, the LAVSM encodes the cross-modal similarity between visual and audio features computed at various temporal offsets, while the AVSD represents the values strictly along its main diagonal to explicitly measure direct temporal alignment.

For authentic videos, the visual and audio signals are naturally well-aligned, resulting in a continuous, bright diagonal in the LAVSM and relatively high AVSD values across all time steps. In easy deepfakes, the audio-visual misalignment is pronounced, causing high response regions in the LAVSM to severely deviate from the diagonal. This deviation visually manifests as scattered bright spots, reflecting the presence of mismatched semantic content or significant temporal shifting. A similar pattern occurs in naive pseudo-fakes, which simply blend random snippets into authentic videos without carefully designed appearance or temporal constraints, thereby introducing obvious, abrupt inconsistencies at the manipulation boundaries. Training a detector on such trivially manipulated samples can easily lead to overfitting and poor generalization to unseen forgeries. In contrast, our AVPF method carefully curates the generation process to produce subtle inconsistencies. As shown in column (d), our method maintains a generally diagonal structure in the LAVSM but introduces localized, fine-grained perturbations that closely mimic the subtle artifacts left by advanced deepfake models. This physically plausible difficulty guides the model to focus on more informative and generalizable temporal cues, rather than memorizing trivial forgery artifacts.

\begin{figure*}[!t]
    \centering
    \includegraphics[width=0.48\linewidth]{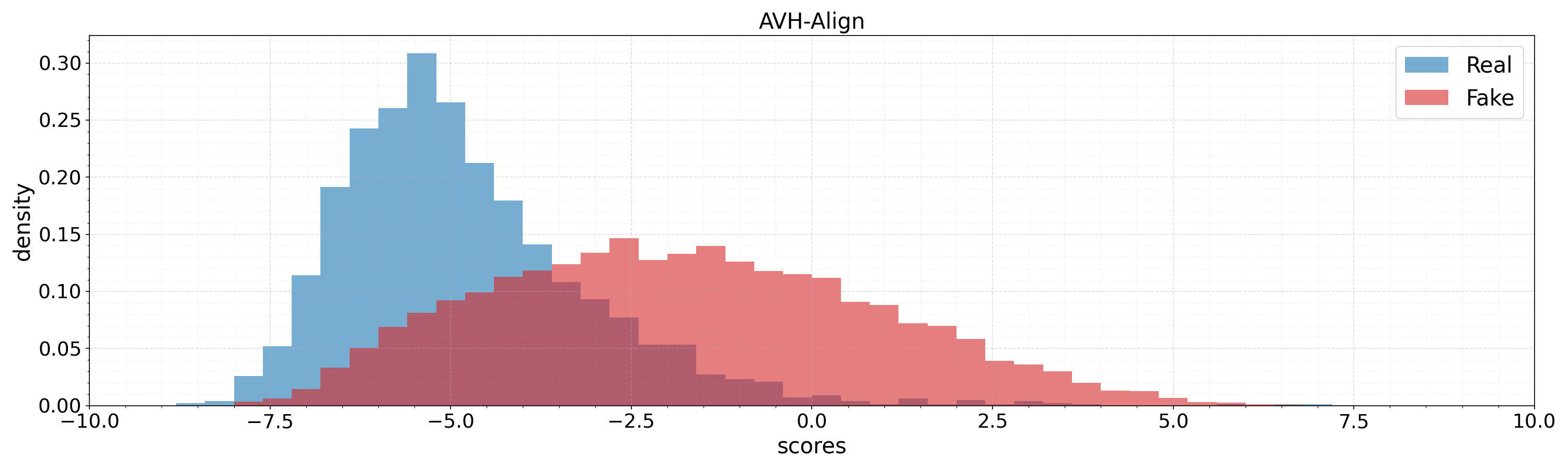}
    \includegraphics[width=0.48\linewidth]{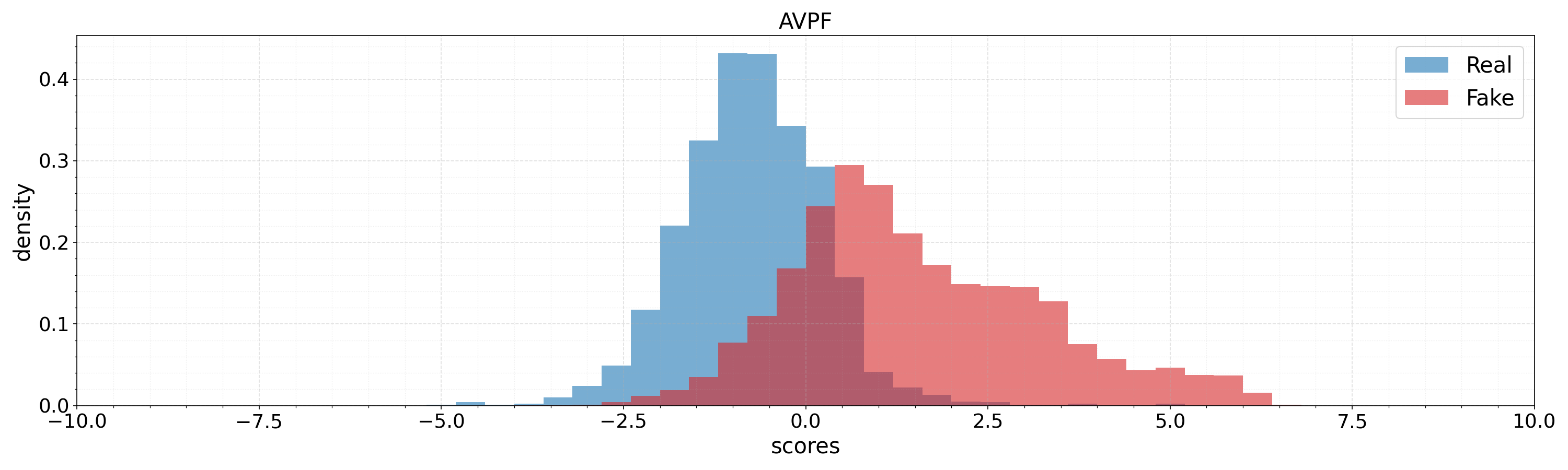} \\
    \includegraphics[width=0.48\linewidth]{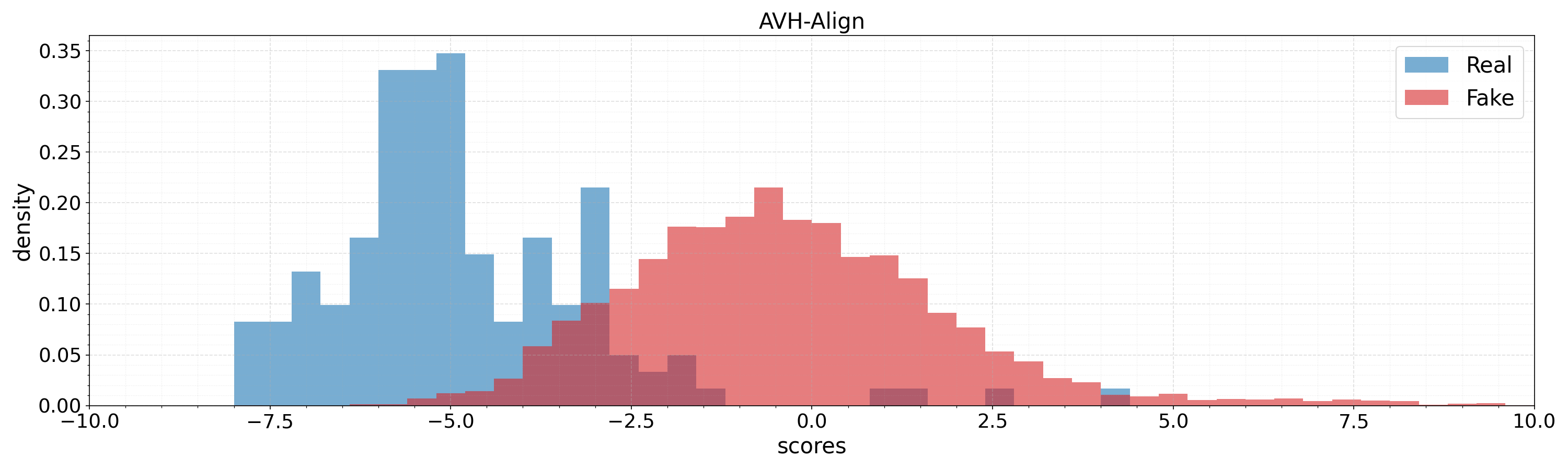}
    \includegraphics[width=0.48\linewidth]{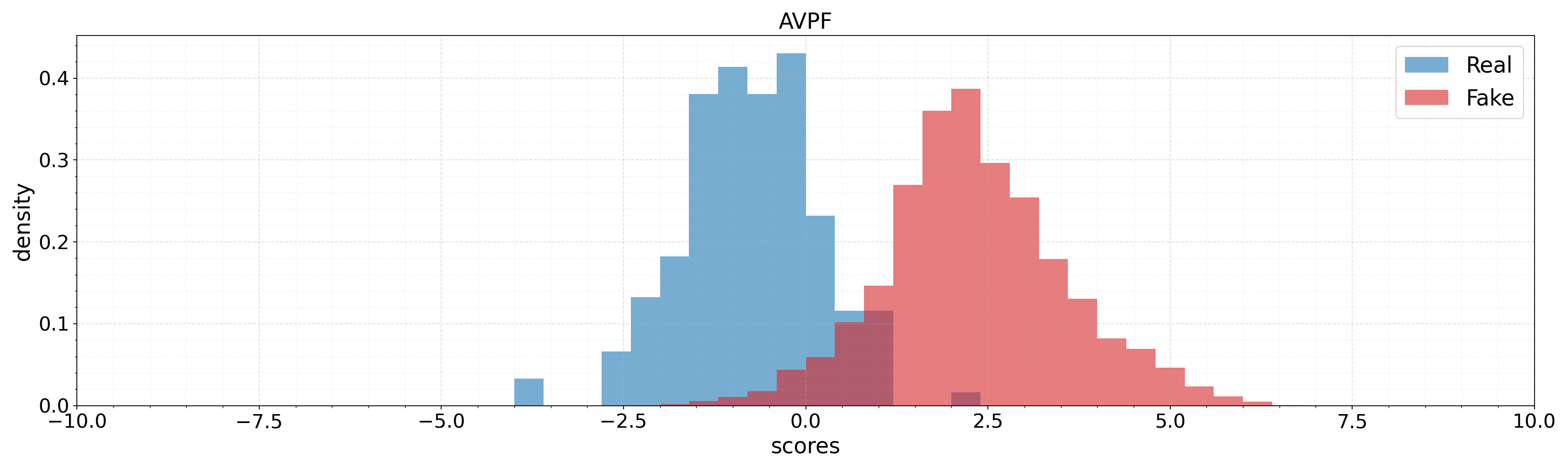} \\
    \includegraphics[width=0.48\linewidth]{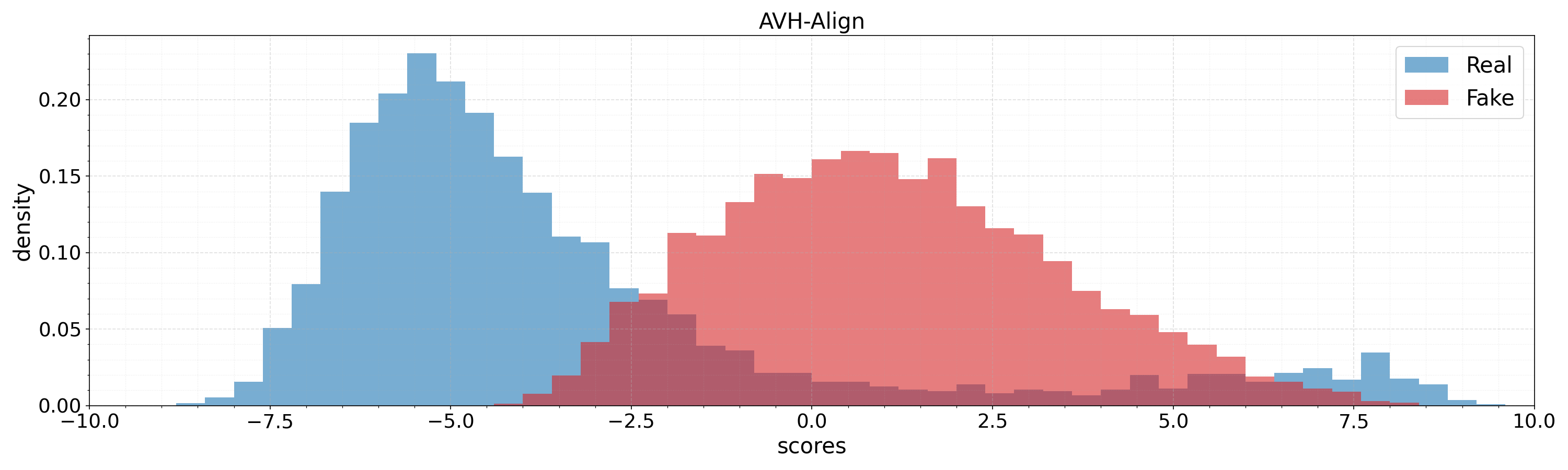}
    \includegraphics[width=0.48\linewidth]{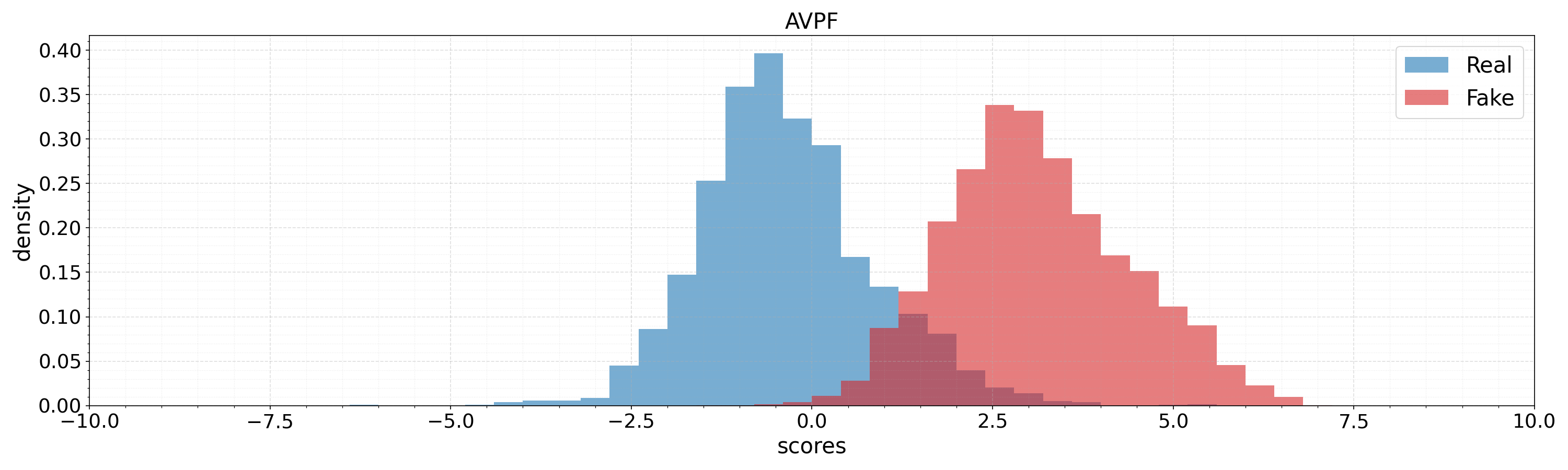} \\
    \vspace{-0.5cm}
    \caption{\small Distribution of prediction scores for AVH-Align (left column) and AVPF (right column) on the AV1M (top), FAVC (middle) and AVLips (bottom) datasets.}
    \Description{Six prediction-score histograms compare AVH-Align and AVPF on AV1M, FakeAVCeleb, and AVLips, showing the distributions of real and fake samples.}
    \label{fig:Distribution_AV1M}
    \vspace{-0.4cm}
\end{figure*}

\begin{figure*}[t]
    
    \centering
    \includegraphics[width=0.195\linewidth]{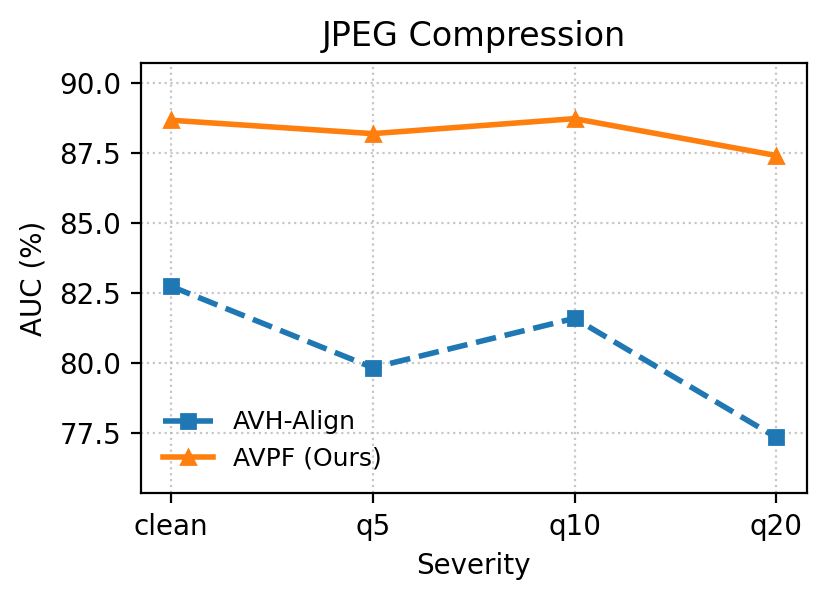} \hfill
    \includegraphics[width=0.195\linewidth]{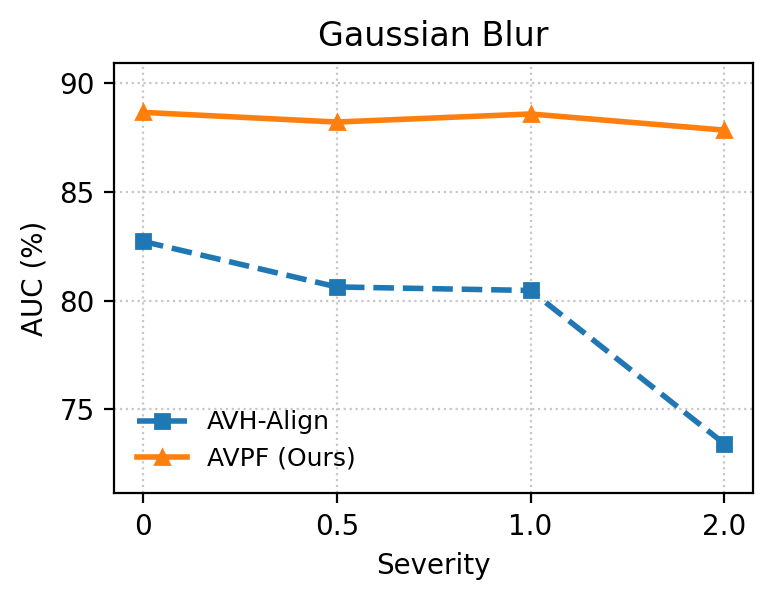} \hfill
    \includegraphics[width=0.195\linewidth]{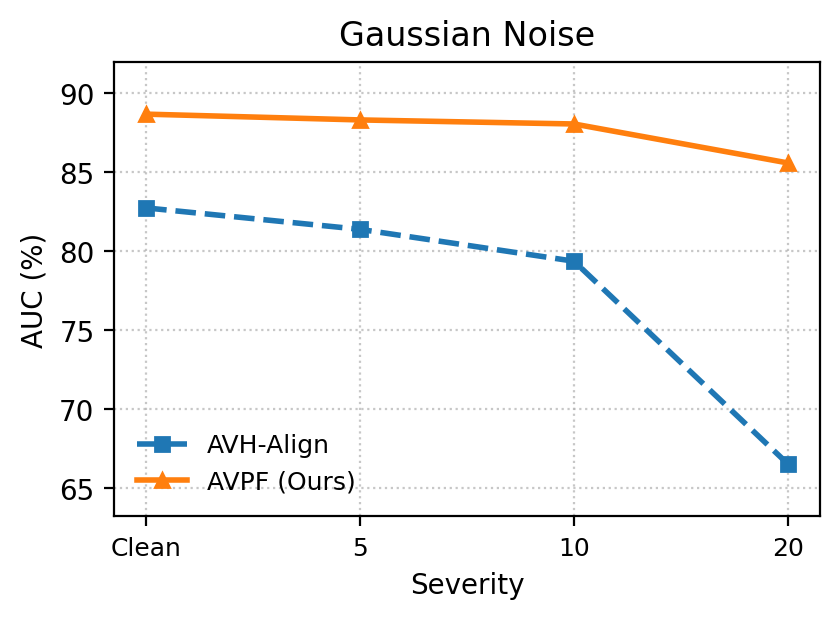} \hfill
    \includegraphics[width=0.195\linewidth]{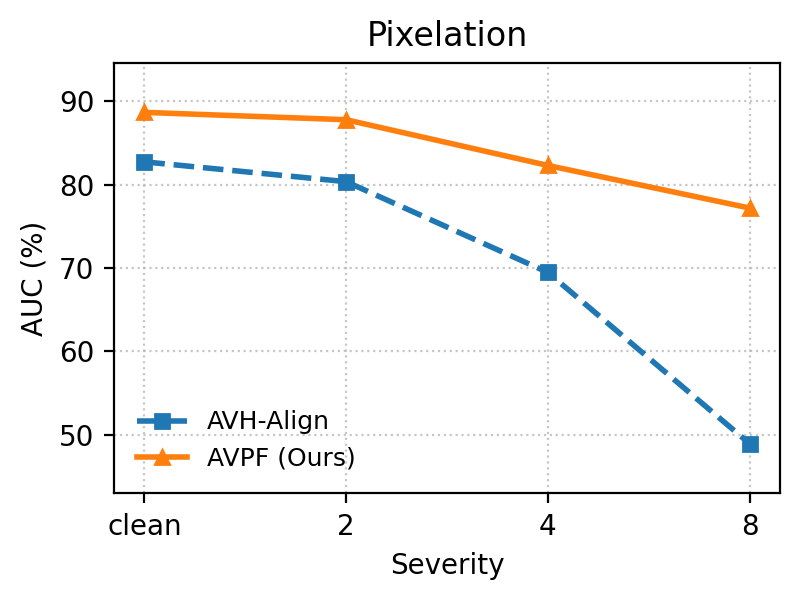} \hfill
    \includegraphics[width=0.195\linewidth]{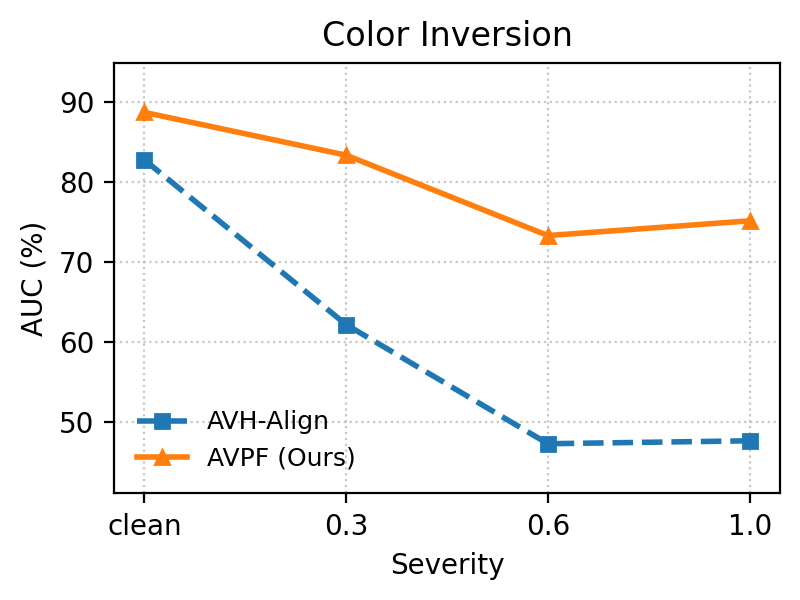} \\
    
    \vspace{-0.1cm} 
    
    \includegraphics[width=0.195\linewidth]{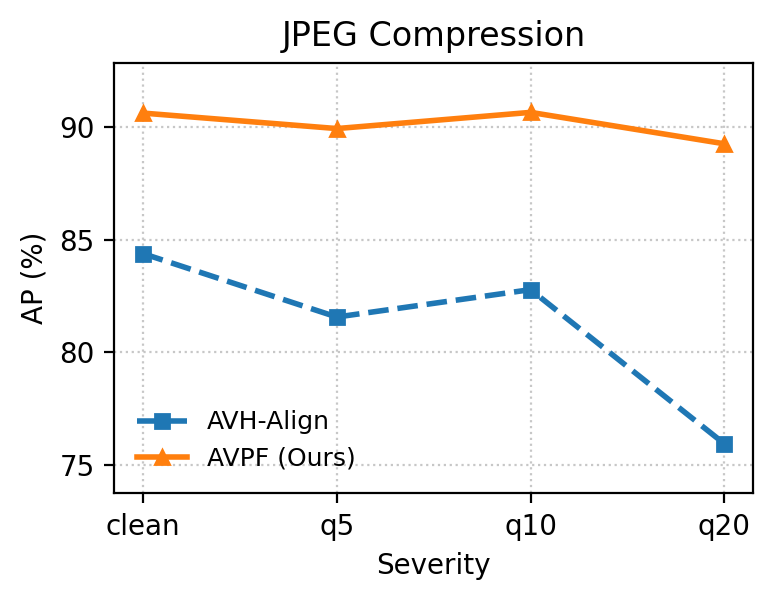} \hfill
    \includegraphics[width=0.195\linewidth]{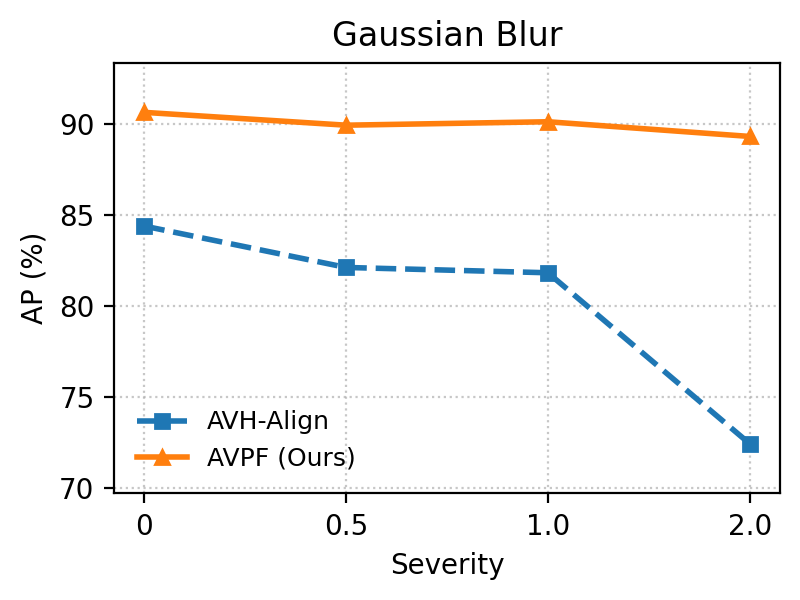} \hfill
    \includegraphics[width=0.195\linewidth]{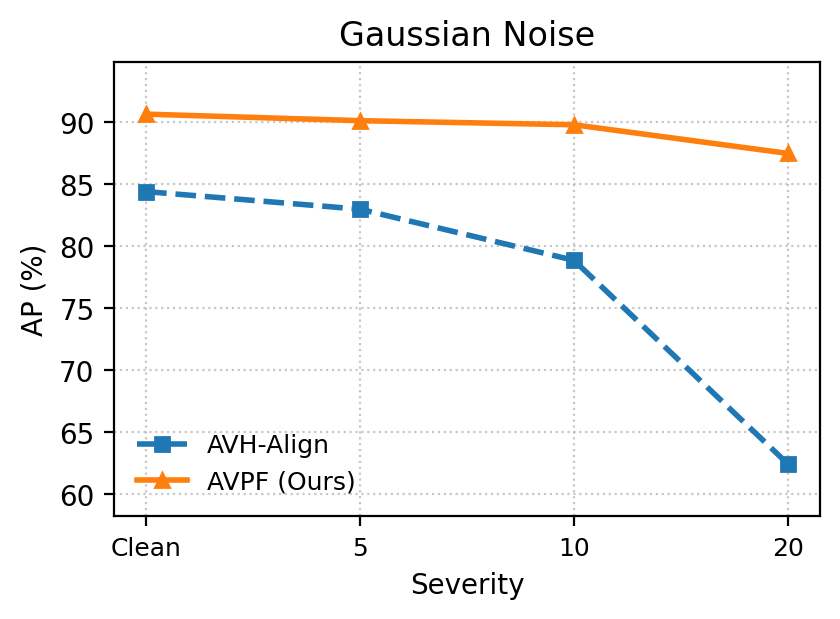} \hfill
    \includegraphics[width=0.195\linewidth]{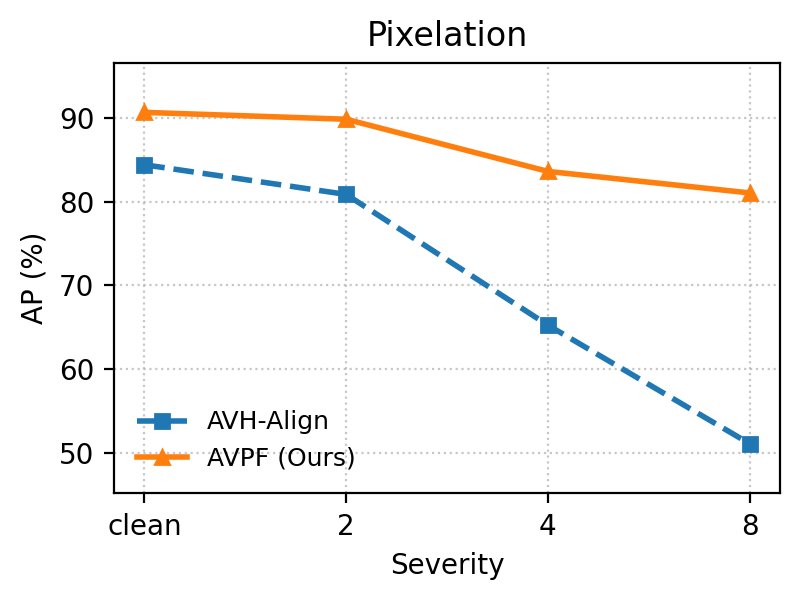} \hfill
    \includegraphics[width=0.195\linewidth]{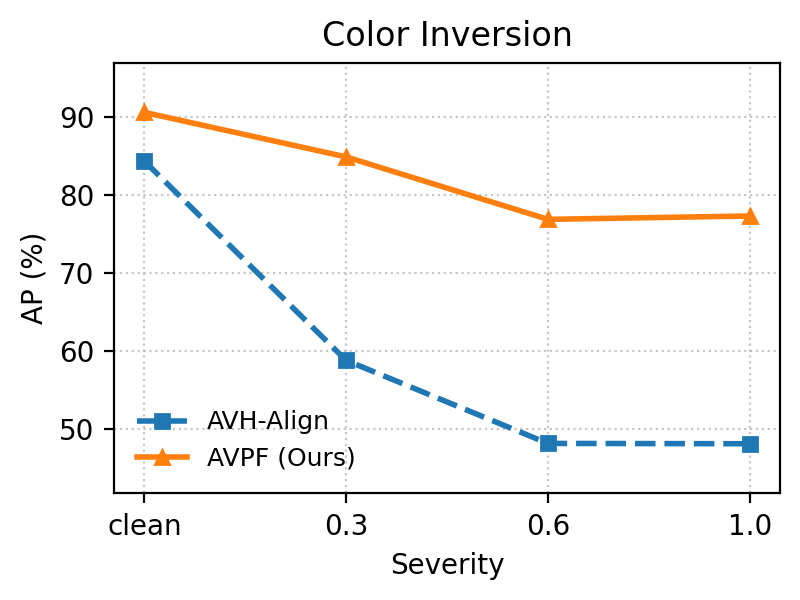}
    \vspace{-0.5cm}
    \caption{\small Robustness comparison between AVH-Align and AVPF (ours) under five different image degradations: JPEG compression, Gaussian blur, Gaussian noise, pixelation, and color inversion (from left to right). The top row displays the AUC curves, while the bottom row shows the AP metrics on the AV1M subset.}
    \Description{Ten line charts compare AVH-Align and AVPF robustness under JPEG compression, Gaussian blur, Gaussian noise, pixelation, and color inversion using AUC and AP.}
    \label{fig:robust}
    \vspace{-0.2cm}
\end{figure*}

\smallskip\noindent\textbf{Prediction Distribution Visualization.}
Figure~\ref{fig:Distribution_AV1M} visualizes the score distributions of real and fake predictions on the test splits of three datasets: AV1M, FAVC, and AVLips. For each dataset, we plot normalized histograms of the model output scores for ground-truth real and fake samples, comparing AVH-Align with our method, AVPF. Across all three datasets, AVPF achieves a better separation between the two distributions, manifested by reduced overlap and a more consistent rightward shift of fake scores relative to real scores. This enlarged margin indicates that AVPF produces more discriminative scoring behavior and supports a more stable decision boundary. In contrast, AVH-Align exhibits heavier overlap and more ambiguous regions, which may lead to increased confusion under a single decision threshold.

    

\smallskip\noindent\textbf{Robustness to Post-processing Operations.}
To evaluate robustness under realistic post-processing conditions, we construct a held-out AV1M test subset consisting of 250 genuine and 250 forged videos, and test AVH-Align and our AVPF method under five common degradations: JPEG compression, Gaussian blur, Gaussian noise, pixelation, and color inversion. As illustrated in Fig.~\ref{fig:robust}, AVPF consistently outperforms AVH-Align across all types and levels of perturbations in both AUC and AP. Although both methods experience performance drops as the severity of corruption increases, AVPF remains significantly more stable, particularly under severe pixelation and color inversion, where AVH-Align suffers drastic declines. These results indicate that AVPF captures more robust forgery cues, facilitating better generalization in practical scenarios.


\smallskip
\noindent\textbf{Limitations and Future Works.} 
Several promising directions remain unexplored: 1) While our method performs well in identifying authenticity, it does not localize where and when the forgery occurs. In the future, we aim to develop strategies capable of both forgery identification and localization. 2) The current study builds upon AV-HuBERT to extract audio-visual representations. Future work will investigate alternative audio-visual backbones and further examine the generalizability of AVPF. 3) We also intend to enrich the pseudo-fake generation process with more sophisticated manipulation operations, including transformations across multiple temporal scales and representation levels. 4) We plan to evaluate the proposed method under more diverse in-the-wild conditions.

\section{Conclusion}
This paper describes a novel self-generated Audio-Visual Pseudo-Fake (AVPF) generation method that notably improves the generalizability of video deepfake detection. The core idea is to produce training data that models the inter- and intra-modality inconsistency commonly observed in real-world video deepfakes. To achieve this, we propose two strategies, Audio-Visual Self-Blending and Audio-Visual Self-Splicing, which effectively create pseudo-fake samples that comprehensively simulate the deepfake data distribution. The proposed method is simple yet effective and relies solely on authentic samples. Experimental results demonstrate its efficacy across various datasets, showing its potential as a strong baseline for multi-modal pseudo-fake-based deepfake detection.

\begin{acks}
This work was supported by the National Natural Science Foundation of China (No.62402464) and the Natural Science
Foundation of Shandong Province (No.ZR2024QF035). 
\end{acks}

\balance
\bibliographystyle{ACM-Reference-Format}
\bibliography{reference}

\appendix
\onecolumn
\section{Supplementary}

\subsection{Visualization of Intra-Modality Inconsistencies}

To intuitively illustrate how our Audio-Visual Self-Splicing (AVSS) strategy simulates real-world deepfakes, we provide a detailed visual and acoustic comparison between authentic videos, real localized deepfakes, and our AVSS pseudo-fakes.

\textbf{Artifacts in Real-World Deepfakes.} In real-world localized deepfakes, manipulators typically edit short, specific video segments (\eg, altering a few words in a sentence). This process inevitably leaves intra-modality temporal traces at the manipulation boundaries. As shown in the visualization of real fake videos, calculating the inter-frame difference ($t$ vs. $t-1$) reveals abnormal, abrupt spatial jumps, highlighted by high-magnitude regions in the difference map. Concurrently, in the audio domain, these unnatural edits cause distinct waveform discontinuities (\eg, sudden amplitude drops or unnatural muting) and abrupt formant shifts or vertical cuts in the Mel-spectrogram.

\textbf{Simulation via AVSS.} Remarkably, our AVSS strategy successfully replicates these exact intra-modal patterns using solely authentic data. By splicing a carefully selected, highly similar short snippet (\eg, 1 frame) into the timeline, AVSS minimizes obvious semantic mismatches while deliberately introducing structural inconsistencies. In the visual domain, the AVSS pseudo-fake exhibits a comparable inter-frame difference at the splicing boundary. Furthermore, the corresponding audio waveform and Mel-spectrogram display sudden phase discontinuities and formant shifts that closely mirror the acoustic artifacts found in actual deepfakes.

\begin{figure}[htbp] 
  \centering
  
  \includegraphics[width=0.95\linewidth]{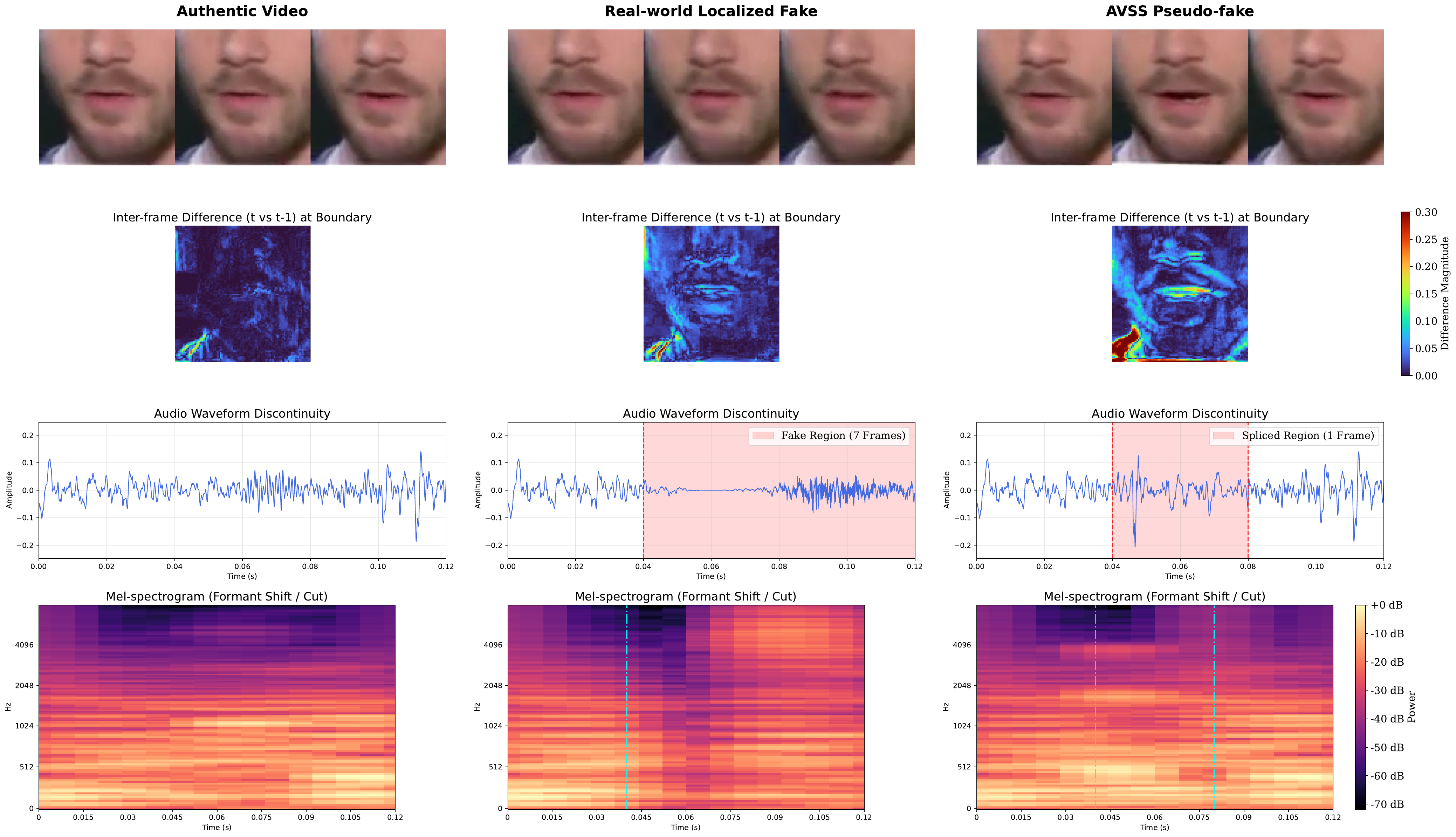}\\
  \vspace{0.1cm}
  \small{(a) Intra-modality inconsistencies in a real-world localized deepfake.}\\

\end{figure}

\textbf{Discussion.} This visualization confirms that AVSS does not simply introduce meaningless noise or trivial artifacts. Instead, it synthesizes physically realistic and meaningful temporal inconsistencies. By training on these self-generated, hard pseudo-fakes, the detector is forced to capture fine-grained, generalized intra-modal forgery traces, rather than overfitting to the spatial flaws of a specific generative model. This intrinsically explains the significant boost in cross-dataset generalizability achieved by our framework.

\subsection{Further Ablation Studies on AVPF Configurations}

\textbf{Effect of similarity metric in AVSS.} In our default Audio-Visual Self-Splicing (AVSS) strategy, we select the most similar snippet by calculating the Mean Absolute Error (MAE) of the visual frames, formulated as $\sum |v_i - v'_{i'}|$. A potential concern is that MAE might be dominated by background or global lighting variations rather than facial dynamics, leading to suboptimal selections when lip movements are subtle. To explicitly avoid this, our MAE is computed exclusively on the tightly cropped lip Regions of Interest (ROIs) (\eg, $96 \times 96$ patches) rather than the full-resolution frames. This spatial constraint naturally filters out irrelevant background noise and ensures the metric is driven purely by mouth movements and local appearance.

To further validate this design, we compare our ROI-based MAE with an alternative geometry-based metric: the $L_2$ distance between facial landmarks. As shown in Table~\ref{tab:ablation_metric}, replacing MAE with landmark-based selection leads to a performance drop, lowering the average AUC/AP from 94.9/98.0 to 93.9/97.5. While landmarks effectively capture structural lip dynamics, they completely discard pixel-level appearance cues (\eg, local illumination, skin texture, and teeth visibility). Ignoring these details during splicing introduces abrupt, unnatural spatial artifacts at the temporal boundaries. These obvious flaws make the pseudo-fakes too ``easy'' for the model to identify, distracting it from learning the fine-grained temporal inconsistencies characteristic of realistic deepfakes. Thus, the ROI-based MAE serves as a robust and superior metric for ensuring both spatial continuity and temporal alignment.

\begin{table}[h]
  \caption{Ablation study of similarity metrics for snippet selection in AVSS.}
  \vspace{-0.2cm}
  \label{tab:ablation_metric}
  \centering
  \small
  \begin{tabular}{l cc cc cc cc}
    \toprule
    \multirow{2}{*}{Method} &
    \multicolumn{2}{c}{FAVC} &
    \multicolumn{2}{c}{AV1M} &
    \multicolumn{2}{c}{AVLips} &
    \multicolumn{2}{c}{Average} \\
    \cmidrule(lr){2-3}\cmidrule(lr){4-5}\cmidrule(lr){6-7}\cmidrule(lr){8-9}
    & AUC & AP & AUC & AP & AUC & AP & AUC & AP \\
    \midrule
    AVSS w/ Landmark
      & 96.8 & \textbf{99.9} & 88.4 & 96.0 & 96.6 & 96.6 & 93.9 & 97.5 \\
    AVSS w/ MAE (Default)
      & \textbf{97.8} & \textbf{99.9} & \textbf{89.2} & \textbf{96.2} & \textbf{97.6} & \textbf{97.8} & \textbf{94.9} & \textbf{98.0} \\
    \bottomrule
  \end{tabular}
\end{table}

\textbf{Sensitivity to temporal window lengths and intervals.} In our pseudo-fake generation, the temporal window length dictates the duration of the localized forgery, while the temporal interval controls the density of these manipulated windows across the video sequence. Table~\ref{tab:ablation_hyper} presents a hyperparameter sensitivity analysis varying these two dimensions. 

Our default configuration (length: $[0.5, 1.5]$s, interval: $1.0$s) achieves the optimal average performance (AUC 94.9, AP 98.0). Deviating from these settings gracefully degrades generalizability. For instance, reducing the window length ($[0.3, 1.0]$s) restricts the temporal context needed for the model to reliably capture cross-modal mismatches, whereas increasing it ($[0.8, 2.0]$s) risks over-perturbing the video and destroying informative natural dynamics. Similarly, adjusting the interval alters the forgery density. A shorter interval ($0.8$s) leads to overly dense manipulations that deviate from typical real-world deepfake distributions, whereas a longer interval with shorter windows (length: $[0.3, 1.0]$s, interval: $1.2$s) produces overly sparse traces, resulting in the lowest average performance (93.5/97.4). This confirms that a balanced temporal configuration is crucial for simulating realistic, high-quality deepfakes.

\begin{table}[h]
  \caption{Ablation study of hyperparameter sensitivity for temporal window length and interval in AVPF generation.}
  \vspace{-0.2cm}
  \label{tab:ablation_hyper}
  \centering
  \small
  \begin{tabular}{l cc cc cc cc}
    \toprule
    \multirow{2}{*}{Temporal Setting} &
    \multicolumn{2}{c}{FAVC} &
    \multicolumn{2}{c}{AV1M} &
    \multicolumn{2}{c}{AVLips} &
    \multicolumn{2}{c}{Average} \\
    \cmidrule(lr){2-3}\cmidrule(lr){4-5}\cmidrule(lr){6-7}\cmidrule(lr){8-9}
    & AUC & AP & AUC & AP & AUC & AP & AUC & AP \\
    \midrule
    Len $[0.3, 1.0]$s, Int $1.0$s 
      & 96.8 & \textbf{99.9} & 88.1 & 95.8 & \textbf{97.8} & \textbf{98.1} & 94.2 & 97.9 \\
    Len $[0.8, 2.0]$s, Int $1.0$s 
      & 97.3 & \textbf{99.9} & 87.6 & 95.7 & 97.7 & 98.0 & 94.2 & 97.9 \\
    Len $[0.5, 1.5]$s, Int $0.8$s 
      & 97.1 & \textbf{99.9} & 88.2 & 95.9 & 97.2 & 97.2 & 94.2 & 97.7 \\
    Len $[0.3, 1.0]$s, Int $1.2$s 
      & 96.0 & \textbf{99.9} & 88.2 & 96.0 & 96.2 & 96.2 & 93.5 & 97.4 \\
    \midrule
    Len $[0.5, 1.5]$s, Int $1.0$s (Def.) 
      & \textbf{97.8} & \textbf{99.9} & \textbf{89.2} & \textbf{96.2} & 97.6 & 97.8 & \textbf{94.9} & \textbf{98.0} \\
    \bottomrule
  \end{tabular}
\end{table}

\subsection{More Attempts}

\textbf{Exploration of frame removal in AVSS.} 
Exploration of frame removal. To further enrich the diversity of intra-modal inconsistencies, we investigate a pseudo-forgery paradigm based on frame removal. This approach synchronously deletes selected video frames and their corresponding audio segments, inducing temporally localized content absence to disrupt alignment. Because naive deletion introduces abrupt temporal discontinuities and visual artifacts, we also apply a temporal smoothing strategy around the removed regions to improve the perceptual naturalness of the forged samples.

As shown in Table~\ref{tab:ablation_edit}, however, incorporating frame removal does not improve detection performance. Rather, it markedly degrades cross-dataset generalization. Across all three configurations employing removal, performance on the challenging AV1M dataset remains consistently poor regardless of the smoothing strategy (AUC: 75.4–76.7, AP: 90.8–91.3), pulling down the overall average. Conversely, omitting the frame removal operation altogether (AVSB+AVSS) yields the best results across both datasets (e.g., Average AUC: 91.1, AP: 97.4), establishing a clear performance advantage. Therefore, we exclude frame removal from our default pipeline.
\begin{table}[h]
  \caption{Ablation study of frame removal strategies on FAVC and AV1M.}
  \vspace{-0.2cm}
  \label{tab:ablation_edit}
  \centering
  \small
  \begin{tabular}{l cc cc cc}
    \toprule
    \multirow{2}{*}{Method} &
    \multicolumn{2}{c}{FAVC} &
    \multicolumn{2}{c}{AV1M} &
    \multicolumn{2}{c}{Average} \\
    \cmidrule(lr){2-3}\cmidrule(lr){4-5}\cmidrule(lr){6-7}
    & AUC & AP & AUC & AP & AUC & AP \\
    \midrule
    AVSB+AVSS+removal (w/ smoothing)  & 91.8 & 99.7 & 76.5 & 91.1 & 84.2 & 95.4 \\
    AVSB+removal (w/ smoothing)       & 94.4 & 99.8 & 76.7 & 91.3 & 85.6 & 95.6 \\
    AVSB+removal (w/o smoothing)      & 95.7 & 99.8 & 75.4 & 90.8 & 85.6 & 95.3 \\
    \midrule 
    AVSB+AVSS (Default)               & \textbf{96.7} & \textbf{99.9} & \textbf{85.4} & \textbf{94.9} & \textbf{91.1} & \textbf{97.4} \\
    \bottomrule
  \end{tabular}
\end{table}

\textbf{Exploration of joint audio-visual tampering.} 
Our default AVSB framework employs single-modality self-blending, manipulating only one modality per video. To investigate whether multi-modal compound manipulations can further increase sample diversity and improve generalization, we explored a more challenging scenario where visual and audio forgeries co-occur within the same video.

As shown in Table~\ref{tab:ablation_joint}, however, joint A/V tampering does not provide a consistent advantage over the single-modality setting. While it yields a marginal AUC improvement on FAVC (97.0 vs.\ 96.9), it noticeably degrades performance on the more challenging AV1M benchmark (AUC: 78.7 vs.\ 79.4, AP: 91.9 vs.\ 92.3). This suggests that concurrently mixing audio and visual manipulations within the same sample may introduce complex distributional shifts that hinder generalization under harder test conditions. Consequently, we retain the single-modality AVSB as our default configuration for its superior stability and cross-dataset performance.

\begin{table}[h]
  \caption{Ablation study of joint audio-visual tampering within the same video on FAVC and AV1M.}
  \vspace{-0.2cm}
  \label{tab:ablation_joint}
  \centering
  \small
  \begin{tabular}{l cc cc cc}
    \toprule
    \multirow{2}{*}{Method} &
    \multicolumn{2}{c}{FAVC} &
    \multicolumn{2}{c}{AV1M} &
    \multicolumn{2}{c}{Average} \\
    \cmidrule(lr){2-3}\cmidrule(lr){4-5}\cmidrule(lr){6-7}
    & AUC & AP & AUC & AP & AUC & AP \\
    \midrule
    AVSB (Joint A/V Tampering) & \textbf{97.0} & \textbf{99.9} & 78.7 & 91.9 & 87.9 & 95.9 \\
    AVSB (Default)             & 96.9 & \textbf{99.9} & \textbf{79.4} & \textbf{92.3} & \textbf{88.2} & \textbf{96.1} \\
    \bottomrule
  \end{tabular}
\end{table}

\textbf{Investigation of spatial random masking.}
While our default Audio Self-Blending (ASB) focuses on temporal intensity scheduling, we further explored its potential from a ``spatial'' perspective by applying a random binary mask to selectively modulate localized audio components. This was intended to enrich sample diversity and bolster the model's discriminative robustness.

Contrary to our expectations, Table~\ref{tab:ablation_randmask} demonstrates that spatial random masking consistently underperforms the baseline. The addition of random masks slightly reduces performance on FAVC and causes a pronounced drop on the challenging AV1M benchmark (e.g., AUC drops from 79.4 to 75.9). These results indicate that spatially random occlusions likely disrupt the fine-grained lip and vocal cues critical for audio-visual consistency modeling, introducing harmful noise rather than informative variability. Thus, we discard spatial masking and maintain the standard AVSB formulation.

\begin{table}[h]
  \caption{Ablation study of spatial random masking in ASB on FAVC and AV1M.}
  \vspace{-0.2cm}
  \label{tab:ablation_randmask}
  \centering
  \small
  \begin{tabular}{l cc cc cc}
    \toprule
    \multirow{2}{*}{Method} &
    \multicolumn{2}{c}{FAVC} &
    \multicolumn{2}{c}{AV1M} &
    \multicolumn{2}{c}{Average} \\
    \cmidrule(lr){2-3}\cmidrule(lr){4-5}\cmidrule(lr){6-7}
    & AUC & AP & AUC & AP & AUC & AP \\
    \midrule
    AVSB (ASB w/ Random Masking) & 95.4 & 99.8 & 75.9 & 90.7 & 85.7 & 95.3 \\
    AVSB (Default)               & \textbf{96.9} & \textbf{99.9} & \textbf{79.4} & \textbf{92.3} & \textbf{88.2} & \textbf{96.1} \\
    \bottomrule
  \end{tabular}
\end{table}


\end{document}